# RASER MRI: Magnetic Resonance Images formed Spontaneously exploiting Cooperative Nonlinear Interaction

**Authors:** Sören Lehmkuhl[1,2]*, Simon Fleischer[3], Lars Lohmann[3], Matthew S. Rosen[4,5], Eduard Y. Chekmenev[6,7], Alina Adams[3], Thomas Theis[2,8,9]*, Stephan Appelt[3,10]*

**Affiliations:**

[1]Institute of Microstructure Technology, Karlsruhe Institute of Technology; 76344 Eggenstein-Leopoldshafen, Germany.

[2]Department of Chemistry, North Carolina State University; Raleigh, NC 27606, USA.

[3]Institute of Technical and Macromolecular Chemistry, RWTH Aachen University; 52056 Aachen, Germany.

[4]Massachusetts General Hospital, A. A. Martinos Center for Biomedical Imaging, Boston, MA 02129, USA.

[5]Department of Physics, Harvard University; Cambridge, MA 02138, USA.

[6]Department of Chemistry, Integrative Biosciences (Ibio), Karmanos Cancer Institute (KCI), Wayne State University Detroit, MI 48202 USA.

[7]Russian Academy of Sciences Leninskiy Prospekt 14, Moscow, 119991 Russia.

[8]Department of Physics, North Carolina State University Raleigh, NC 27695 USA.

[9]Joint Department of Biomedical Engineering, University of North Carolina at Chapel Hill & North Carolina State University, Raleigh, NC 27695 USA.

[10]Central Institute for Engineering, Electronics and Analytics – Electronic Systems (ZEA-2), Forschungszentrum Jülich GmbH, D-52425 Jülich, Germany.

*Corresponding author. Email: lehmkuhl@kit.edu, ttheis@ncsu.edu, st.appelt@fz-juelich.de

**Abstract:** The spatial resolution of magnetic resonance imaging (MRI) is fundamentally limited by the width of Lorentzian point spread functions (PSF) associated with the exponential decay rate of transverse magnetization ($1/T_2^*$). Here we show a different contrast mechanism in MRI by establishing RASER (Radio-frequency Amplification by Stimulated Emission of Radiation) in imaged media. RASER imaging bursts emerge out of noise and without applying (Radio Frequency) RF pulses when placing spins with sufficient population inversion in a weak magnetic field gradient. A small difference in initial population inversion density creates a stronger image contrast than conventional MRI. This contrast is based on the cooperative nonlinear interaction between all slices. On the other hand, the cooperative nonlinear interaction gives rise to imaging artifacts, such as amplitude distortions and side lobes outside of the imaging domain. Both the contrast and the artifacts are demonstrated experimentally and predicted by simulations based on a proposed theory. This theory of RASER MRI is strongly connected to many other distinct fields related to synergetics and non-linear dynamics.

**One-Sentence Summary:** Spontaneous collective emission of radiofrequency bursts from nuclear spins in the presence of a gradient enables a new form of magnetic resonance imaging.

**Introduction**

RASER (Radio-frequency amplification of stimulated emission of radiation), also referred to as Zeeman MASER, is a Nuclear Magnetic Resonance (NMR) phenomenon as a result of stimulated nuclear spin transitions. RASERs have been investigated using hyperpolarized rare-gases (*1-4*) as well as $^1$H, $^{17}$O and even $^{27}$Al spins in liquids and solids (*5-9*). Multimode RASERs enable co-magnetometry, which in turn allows for precision measurements (*10-13*). Additionally, multimode RASER activity gives insight into fundamental phenomena in nonlinear mathematics (*14*) and synergetics (*15*) such as line collapse, multiple period doubling, intermittence and chaos(*4, 12, 16*). Most recently, the parahydrogen (*p*-H$_2$) pumped (*17, 18*) RASER has been established (*12, 16, 19-21*), by creating strong population inversions directly in room temperature solutions. The *p*-H$_2$ RASER is associated with significantly extended decoherence times and it appears natural to wonder whether it could serve as a means to overcome fundamental limits of MRI (*22, 23*).

The spatial resolution of magnetic resonance imaging (MRI) is limited by the width w = $1/(\pi T_2^*)$ of the Lorentzian point spread function (PSF). Here we show that nonlinearly coupled slices can spontaneously form an image out of spin noise, as an alternative to the superposition of uncoupled Lorentzian PSF's. We describe novel nonlinear MRI physics in a *p*-H$_2$ RASER, while noting that nonlinear spin evolution in the presence of a gradient including radiation damping effects and dipolar fields has been reported before (*24, 25*). We note that other hyperpolarization techniques can also be also employed for RASER MRI described here. (*7*)

Conventional MRI uses spin or gradient echoes of nuclear magnetization that need to be excited with RF pulses. An interesting alternative is spin noise imaging, which measures projections without RF excitation and fast gradient switching, but is inherently less sensitive (*26*). Thus spin noise imaging requires cryogenically cooled NMR probes and long averaging times to compensate for the low signal to noise ratio (SNR).

Instead, the system under study here achieves an SNR > 200 at room temperature in a single scan by emitting spontaneous RASER bursts without RF excitation. In the time domain, these bursts reflect the superposition of nonlinearly coupled slices. The corresponding spectra of the bursts report on the spatial distribution of the object. Because of the nonlinear coupling, the spectra can have complicated and distorted shapes. So, despite the advantage of better image contrast, RASER MRI entails new MRI physics challenges and opportunities caused by the nonlinear coupling.

**RASER MRI Theory**

In the presented work, RASERs emerge when placing a proton spin 1/2 ensemble with a large initial population inversion, $d_0 = N_2-N_1$, above the RASER threshold $d_{th} = 4V_s/(\mu_0 \hbar \gamma_H^2 T_2^* Q)$ in a resonant LC circuit with quality factor $Q$. $N_2$ and $N_2$ are the populations of the corresponding Zeeman levels 2 and 1, $V_s$ is the sample volume, $\mu_0, \hbar$ and $\gamma_H$ denote the vacuum permeability, Planck's constant and the proton gyromagnetic ratio, respectively. For RASER MRI the proton spins are first pumped into a state of highly negative spin polarization $P_H$. This corresponds to a positive $d_0$ value, which is assumed to be several orders of magnitude above the RASER threshold $d_{th}$. An equivalent and convenient way to characterize the threshold condition for one singular mode is given by $\varepsilon = d_0/d_{th} = T_2^*/\tau_{rd} \gg 1$ (*27, 28*), where $\varepsilon$ is a dimensionless quantity. Note that epsilon is the enhancement above the RASER threshold, not above thermal nuclear spin polarization. The radiation damping rate is given by $\tau_{rd}^{-1} = \mu_0 \hbar \gamma_H^2 Q\, d_0/(4V_s)$, which includes inverted states (positive $d_0$), and it has been studied extensively in NMR spectroscopy (*24, 27, 29-31*).

In order to understand how the RASER can be utilized for MR imaging, we introduce an analysis of the RASER action in the presence of a magnetic field gradient $G_z$. The gradient creates a frequency range $\Delta = \gamma_H \cdot G_z \cdot L$ that spans the image domain of the object of length L (SI, section 1). The initial nuclear spin population inversion is spread over the imaging domain $\Delta$ and is given by $d_0 = \int_{\nu_0 - \frac{\Delta}{2}}^{\nu_0 + \frac{\Delta}{2}} \rho_d(\nu) d\nu$, where $\rho_d(\nu)$ is the population inversion density and $\nu_0$ is the off-resonance frequency in the center of the imaging domain. The integrand $\rho_d(\nu) d\nu$ can be described as the number of negatively polarized spins in the frequency interval $[\nu, \nu+d\nu]$. Given a profile $\rho_d(\nu)$, a total RASER MRI signal emerges spontaneously out of the spin noise.

To generate a system where a numerical evaluation is feasible, we divide the image domain $\Delta$ into $N = \Delta/\delta\nu$ individual slices. To avoid numerical artifacts, the distance $\delta\nu$ between consecutive slices has to be chosen small enough. Specifically $\delta\nu < w$ has to be fulfilled, where $w = 1/(\pi T_2^*)$ is the natural linewidth. Furthermore, to estimate whether a given $d_0$ is RASER active in a given gradient $G_z$, we also introduce the threshold population density $\rho_d^{th} = d_{th}/w$ as used below.

To calculate the dynamics of the nonlinearly coupled slices, each slice $\mu=1\ldots N$ is characterized by an initial population inversion $d_\mu(0) = \int_{\nu_0 - \frac{\Delta}{2} + (\mu-1)\delta\nu}^{\nu_0 - \frac{\Delta}{2} + (\mu)\delta\nu} \rho_d(\nu) d\nu$. With a given initial $d_\mu(0)$, the time evolution of the RASER modes or slices can be modeled by a set of $\mu = 1..N$ nonlinear coupled differential equations for the population inversion $d_\mu$ and the transverse spin component $\alpha_\mu = A_\mu \exp(i\phi_\mu)$.

$$\dot{d}_\mu = -\frac{d_\mu}{T_1} - 4\beta \sum_{\sigma,\tau=1}^{N} A_\sigma A_\tau \cos(\phi_\sigma - \phi_\tau) \qquad (1),$$

$$\dot{A}_\mu = -\frac{A_\mu}{T_2^*} + \beta d_\mu \sum_{\tau=1}^{N} A_\tau \cos(\phi_\tau - \phi_\mu) \qquad (2),$$

$$\dot{\phi}_\mu = 2\pi\{\nu_0 - 0.5[\Delta - \delta\nu(2\mu-1)]\} + \beta \frac{d_\mu}{A_\mu} \sum_{\tau=1}^{N} A_\tau \sin(\phi_\tau - \phi_\mu) \qquad (3),$$

$$d_\mu(0) = \int_{\nu_0 - \Delta/2 + (\mu-1)\delta\nu}^{\nu_0 - \Delta/2 + \mu\delta\nu} \rho_d(\nu) d\nu \qquad (4).$$

The coupling constant $\beta$ is given as $\beta = \mu_0 \hbar \gamma_H^2 Q/(4V_s)$. The model for RASER MRI represented by Eqs.(1-4) is formulated in the rotating frame (for a complete derivation see SI, section 1), and is a modification of the existing multi-mode RASER theory (*12, 16*). The modifications comprise the initial boundary conditions for $d_\mu(0)$ in Eq.(4), the absence of pumping in Eq.(1) and the definition of the slice frequencies in Eq.(3). We assume a random fluctuation (nuclear spin noise excitation) for the initial small amplitudes $A_\mu(0)$ and phases $\phi_\mu(0)$, which initiate the self-induced RASER burst in the absence of any RF excitation. Numerical simulations of Eqs.(1-4) reveal three important invariance principles for RASER MRI: Provided that $\delta\nu < w$ and $T_1 \gg T_2^*$, the amplitude and contrast properties of the RASER images are independent of (I) the value of the slicing $\delta v$, (II) the longitudinal relaxation time $T_1$ and (III) the values of the initial conditions $A_\mu(0)$ and $\phi_\mu(0)$.

Certain processes can be identified by examining the dynamics described by Eqs (1-3): The population inversion of a given mode $\mu$ in Eq.(1) decays with the rate $1/T_1$ and is decreased by the rate given by the sum over all quadratic terms $-4\beta A_\sigma A_\tau \cos(\phi_\sigma - \phi_\tau)$. In turn, the amplitude of $A_\mu$ in Eq.(2) decays with the rate $1/T_2^*$ and increases for $\tau = \mu$ with the rate $\beta \cdot d_\mu$. The last term on the right side of Eq.(2), $\beta d_\mu \sum_{\tau=1}^{N} A_\tau \cos(\phi_\tau - \phi_\mu)$ for $\tau \neq \mu$ involves a sum over all other amplitudes $A_\tau \cos(\phi_\tau - \phi_\mu)$. This sum can be a growth or decay rate for $A_\mu$, depending on the specific values of all other phase differences $\phi_\tau - \phi_\mu$. The collective action of all modes strongly influences the amplitude and sign of the rate $dA_\mu/dt$, which defines the amplitudes $A_\mu$ of the final image.

The spatial encoding of each slice $\mu = 1..N$ is reflected by the first term in Eq.(3), where each slice is oscillating at the angular frequency $\omega_\mu = 2\pi(\nu_0 - 0.5(\Delta - \delta v(2\mu-1)))$. Apart from this linear evolution of $\phi_\mu$ with time $t$, there is a nonlinear collective term $(\beta d_\mu/A_\mu) \sum_{\tau=1}^{N} A_\tau \sin(\phi_\tau - \phi_\mu)$ which is responsible for synchronism. In fact Eq.(3) is analogue to Kuramoto's model of synchronized oscillators (*32-34*). The dynamics of RASER MRI given by Eqs.(1-4) can be described by a collection of synchronized oscillators or slices with distinct angular frequencies $\omega_\mu$, where the amplitude $A_\mu$ of each oscillator depends on the self-organization controlled by the collective interaction with all other slices. Therefore, the derivative of the amplitude of each slice depends on the mean-field amplitude produced by all other slices.

Finally, the total RASER signal is obtained by the sum of all transverse spin components $Sig(t) = \sum_{\mu=1}^{N} \text{Re}(\alpha_\mu) = \sum_{\mu=1}^{N} A_\mu \text{Re}(\exp[i\phi_\mu])$. Here, we focus on the difference between the concept of single PSF's to analyze conventional MR image formation. and the collective mean-field approach, which is the basis of RASER MRI. Numerical solutions of Eqs.(1-4) are evaluated (see Fig.1) in order to highlight the difference of the spin dynamics for a single RASER slice and the collective behavior of coupled slices.

The simplest case is shown in Fig. 1(A) for $N = 1$ and $T_1 = \infty$, where the numerically evaluated form matches the exact solution introduced by Mao *et al.* (*29, 35, 36*) also discussed by others (*37, 38*). The corresponding phased and absolute spectrum of $\alpha = \alpha_1$ are displayed on the bottom right of Fig.1(A). For this case $T_1 = \infty$, the PSF is a hyperbolic secant with width $w_{sech}$ (SI, section 2, Eq. S19). Close to the threshold, such a PSF is narrower than the Lorentzian NMR linewidth $w = 1/(\pi T_2^*)$, because the RASER signal involves dedamping.

No exact solution exists for a finite $T_1$, but the MR signal represents an asymmetrically shaped PSF (Fig. 1(B), see SI, section 3). The linewidth $w_{as}$ in the spectrum is slightly broader compared to the symmetric case (Fig.1(A)), but still smaller than $w$.

Here, for the first time, we include both the effects of finite $T_1$ and the nonlinear interactions between $N$ slices formed in the presence of a gradient. In contrast to standard MRI, the image contrast and the spatial resolution cannot be explained by independent individual PSFs. Each slice is sensitive to the collective action of all other slices, which makes RASER imaging highly sensitive to local variations in $d_\mu$ [SI, section 4(a)] providing new frontiers in MRI spin physics.

# RASER MRI explored by numerical simulation

In the simulation in Fig. 1(C), a rectangular polarization profile (inset upper right) is assumed to generate a RASER signal in the presence of a field gradient. The time evolution of three of the $N = 30$ slices is depicted on the left. The shape of signal of these slices differs significantly from the uncoupled PSFs in (A) and (B). A corresponding 1D RASER image (projection) is obtained as the Fourier transform from $Sig(t) = \text{Re}(\sum \alpha_\mu)$. The amplitude in the center of the RASER image is larger and decaying side lobes arise outside of the image boundaries at $x = \pm 4$ mm (bottom right). These artifacts are expected from the theory described in Eqs.(1-4) and evaluated in detail by numerical simulations in the SI, section 4.

In Fig. 1(D), we simulate a RASER image using a spin density profile $\rho_d(v)$ to match the experimental setup described below in Fig. 2(A,B). This non-uniform spin density profile $\rho_d(v)$ entails two equal compartments separated by a gap. The evolution of five representative RASER slices out of $N = 50$ coupled slices is shown (Fig. 1(D), left). The image after Fourier transformation (bottom right) reflects roughly the shape $\rho_d(v)$ except for the deformed amplitudes of the flat tops and the side lobes, which occur outside the imaging boundaries.

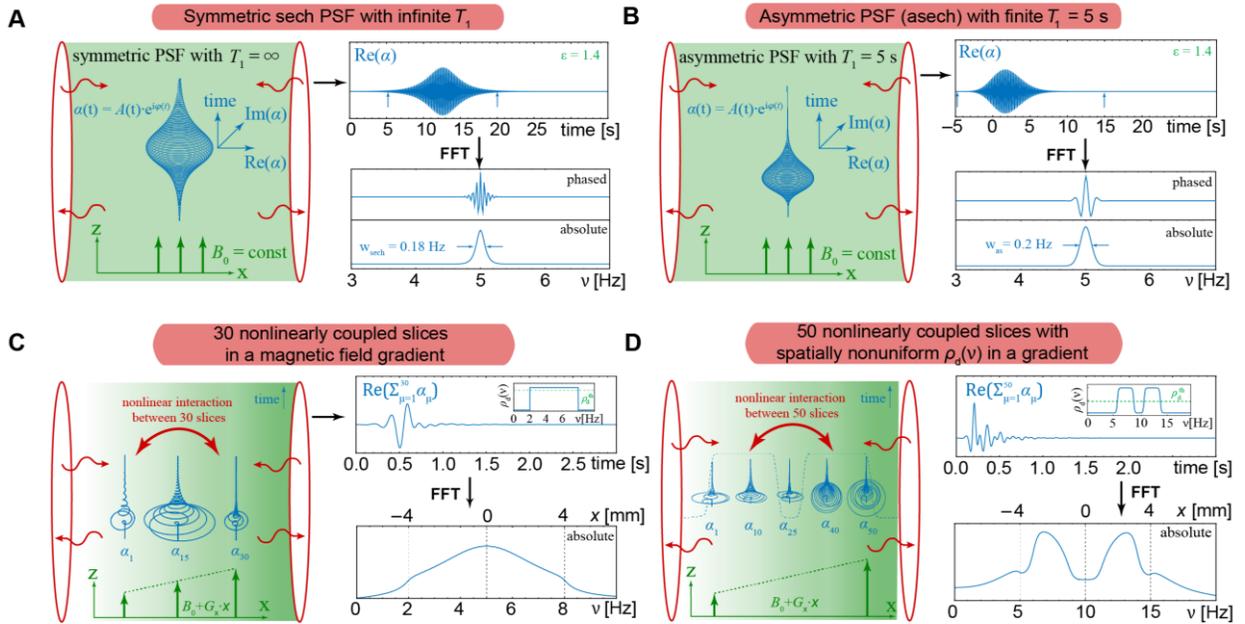

**Fig.1. Simulated RASER signals and the corresponding Fourier transformed spectra for different numbers of interacting slices.** The nonlinear interaction between all slices is mediated by the virtual photons (wavy arrows, wavelength ≫ sample dimension) in the resonator (red) (*39*). After the RASER burst, the Zeeman energy of the spins is fully transferred to the current of the coil (*40*). (**A**) For $N = 1$ and $T_1 = \infty$, the signal $\alpha_1(t) = \alpha(t)$ is plotted in the $(t, \text{Re}(\alpha), \text{Im}(\alpha))$ space (left). The projection $\text{Re}(\alpha)$ for $d_0 = 4.2 \cdot 10^{15}$ and the corresponding Fourier transformed spectra are shown on the right. (**B**) For $N = 1$, $T_1 = 5$ s and $d_0 = 4.2 \cdot 10^{15}$, the signal burst $\alpha(t)$ is asymmetric with respect to time. (**C**) Sketch of three representative signals $\alpha_\mu$, $\mu = 1, 15, 30$ out of $N = 30$ interacting slices ($T_1 = 5$ s, $\Delta = 6$ Hz, rectangular profile with $\rho_d(v) = 7.5 \cdot 10^{15}$/Hz). (**D**) Five representative signals $\alpha_\mu$ out of $N = 50$ coupled slices ($T_1 = 5$ s, $\Delta = 10$ Hz, non-uniform density $\rho_d(v)$). Threshold population density $\rho_d^{th} = d_{th}/w = 6.6 \cdot 10^{15}$/Hz is indicated as dotted line in the insets in (C,D).

**Experimental Realization of RASER MRI: 1D demonstrations**

To experimentally examine the RASER MRI theory, a simple phantom was prepared consisting of a cylindrical sample chamber divided into two measurement chambers by a glass slide (Fig. 2(A,B)). The two chambers are individually supplied with $p$-H$_2$ to generate highly negative polarized proton spins (i.e. $d_0 \gg d_{th}$). The chemical system chosen is pyrazine in a liquid methanol-d$_4$ solution with a dissolved iridium-based SABRE catalyst for nuclear spin polarization (*18, 41*). RASER MR images were acquired in the presence of weak $G_x$ and $G_z$ magnetic field gradients on the order of a few mG/cm.

Conventional MR images (SEI) were obtained with a 90°-180° Spin Echo sequence (Fig. 2(C)) as a reference. Prior to the acquisition of the reference Spin Echo Image (SEI), a crusher field gradient was applied to the hyperpolarized sample, to suppress spontaneous RASER build-up. 1D images were acquired using the $G_z$ gradient in order to visualize the two chambers separated by the dividing glass slide. 2D images were recorded through stepwise switching of the $G_x$ and $G_z$ gradients to rotate through a circle with constant absolute gradient ($|G| = (G_z^2 + G_x^2)^{1/2}$. The 2D image was then obtained via projection reconstruction, which is also common in Computed Tomography (CT).

The RASER images were acquired in a similar way (Fig. 2D), but in contrast to the spin echo sequence no RF-pulses were applied. The signal is acquired in the presence of $G_x$ and $G_z$ field gradients during spontaneous RASER emission, which begins spontaneously, shortly after the crusher field gradient is turned off.

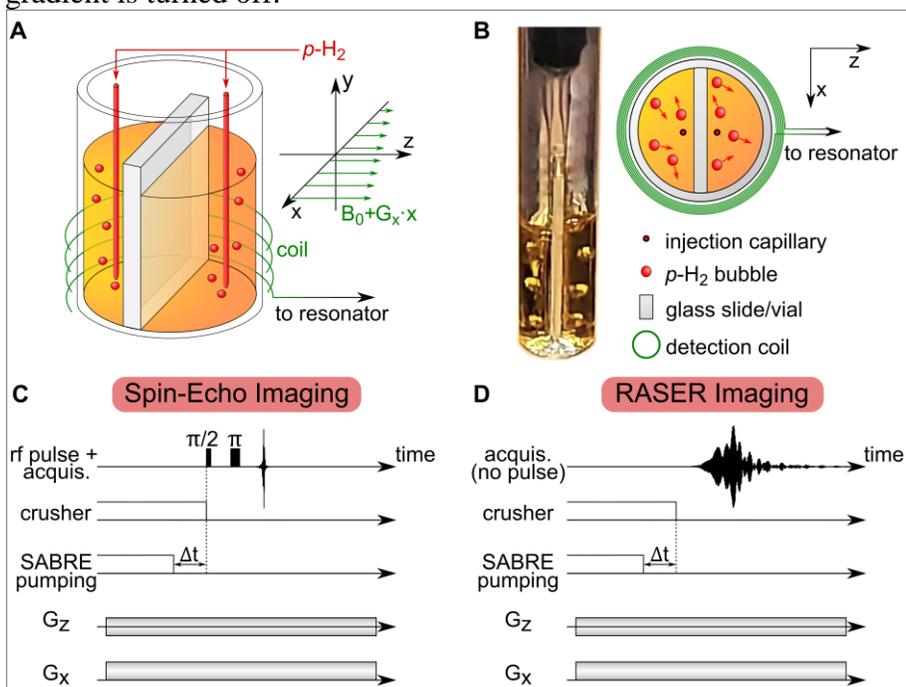

**Fig. 2. Experimental setup for MRI of a two-chamber phantom (top) and corresponding pulse sequences for Spin Echo Imaging (SEI) and RASER MRI (bottom).** (**A**) Schematics of the two imaged chambers and of the gradient directions. (**B**) Photo and top-down schematic of the two chambers (L = 8 mm diameter, 10 mm height separated by a 1 mm thick glass slide) including bubbling of $p$-H$_2$ through two capillaries (100 μm OD and 30 μm ID). (**C**) 90°-180° Spin Echo sequence for SEI. (**D**) RASER imaging sequence. For both imaging sequences a crusher gradient is applied to destroy all coherence while negative proton polarization is built up by SABRE pumping at magnetic fields $B_0$ of 3.9 and 7.8 mT. $p$-H$_2$ bubbling is interrupted to allow the solution to settle for a time $\Delta t$. For SEI, the image is encoded in the echo signal. In the case of RASER MRI the signal builds up spontaneously in the absence of any RF-excitation. Frequency encoding is performed in x and z direction.

The RASER action can be measured over an indefinite period (Fig.3(A)), when $p$-$H_2$ is continuously bubbled through the solution. However, a challenge for imaging under bubbling conditions in the presence of field gradient results from sample motion induced by the bubbling. This collapses the RASER spectrum in each chamber into one average frequency (Fig. 3(B)). In order to avoid line-collapse induced by sample motion, and to enable imaging, the $p$-$H_2$ flow had to be stopped and an additional waiting time $\Delta t$ was introduced, which allows for the solution to settle and the motions to halt. Now, both SE and 1D RASER signals could be acquired (Figs.3(D,G)) shortly after the crusher gradient was switched off. The acquired RASER burst in Fig.3(G) is significantly longer than the corresponding spin echo in (D) acquired at the same gradient strength of $G_z = 3.84$ mG/cm.

The spatial resolution limit is given by $\delta z = w/(\gamma_H G_z)$ in conventional MRI (*22*). This limit yields $\delta z_{SEI} = 280$ μm for the SEI in Fig.3(E), and as a result the gap and the edges of the sample are not well resolved. However, for RASER 1D projection in Fig.3(H), the slope at the image boundaries at the gap is more than three times higher. Thus, a spatial resolution of $\delta z_{RI} \approx 90$ μm is estimated.

The measured 1D RASER image in Fig. 3(H) shows signal lobes outside the boundaries of $z = \pm 4$ mm, in accord with the simulation shown in Fig. 1(D). Such artifacts from 1D RASER MRI are analyzed in section 4(b) of the SI and a potential correction method is proposed.

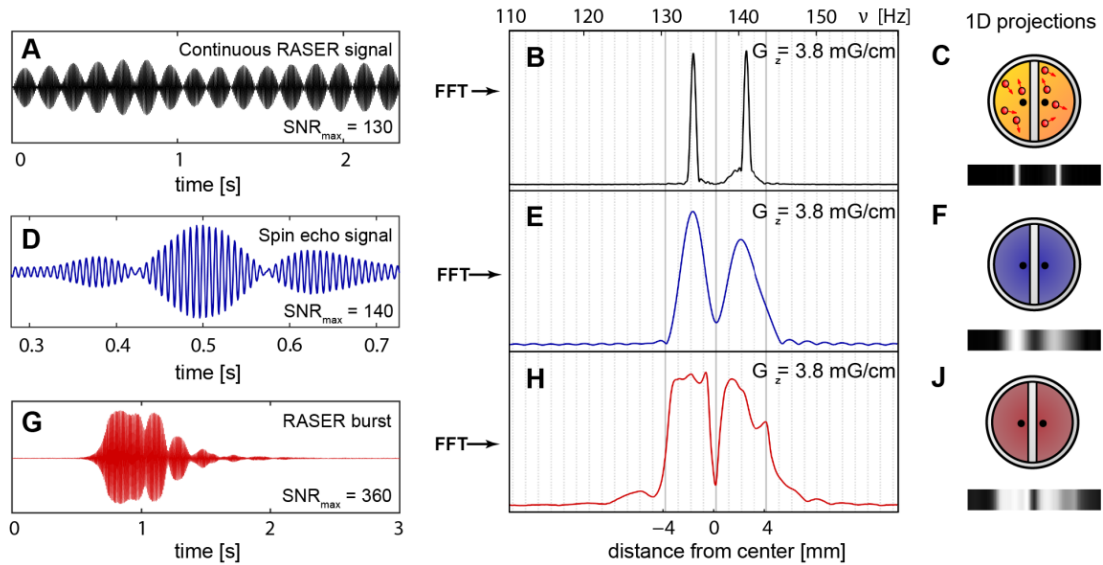

**Fig. 3. 1D projections of a continuously pumped proton RASER, a SEI and a RASER image.** (**A**) Continuously SABRE pumped proton RASER signal and corresponding Fast Fourier Transformed (FFT) spectrum (**B**) in the presence of a gradient $G_z$. A Hamming window is applied to the signal before FFT to suppress Sinc wiggles. (**D**) Spin echo acquired with the sequence in Fig.2(C) and (**E**) corresponding Fourier transformed SEI. (**G**) RASER burst acquired with the sequence in Fig.2D. (**H**) Corresponding RASER 1D projection, which is three times better resolved ($\delta z_{RI} \approx 90$ μm) than the SEI in (E). $B_0 = 7.8$ mT (proton resonance frequency 333 kHz) and no slice selection is applied. The RASER image (H) has $SNR_{max} = 360$ at $\Delta t = 2$ s, while the SEI in (B) yields $SNR_{max} = 140$ at $\Delta t = 5$ s. All images are phased in the absolute mode and were measured in a single scan. (**C,F,J**) Corresponding image phantom and 1D projections.

**Experimental realization of RASER MRI: 2D demonstration and comparison to traditional spin echo image (SEI) of hyperpolarized solutions.**

Both a 2D-SEI (Fig.4(A)) and a 2D-RASER MRI (Fig.4(B)) of the same sample are obtained, extending 1D imaging to 2D imaging by reconstructing from 30 angular directions. The field gradient used for the SEI image was 3.5 times larger than that for RASER MRI in order to obtain comparable resolution. Each individual projection in the SEI has a resolution of 50 μm, only about an order of magnitude higher than modern micro imaging (*42-44*). The two semicircle-shaped halves and the 1-mm gap are clearly visible in both images of Fig.4. These images also display typical projection reconstruction star artifacts outside of the imaging domain. The 2D-RASER image in Fig.4(B) shows sharper features, but also exhibits a deformed shape of the sample and its gap, paired with several interfering lines. These lines are probably caused by residual motion of the liquid after turning off the *p*H$_2$-pumping. They can be identified in the individual 1D projections, which are used to reconstruct the 2D RASER image (see SI, Fig. S11).

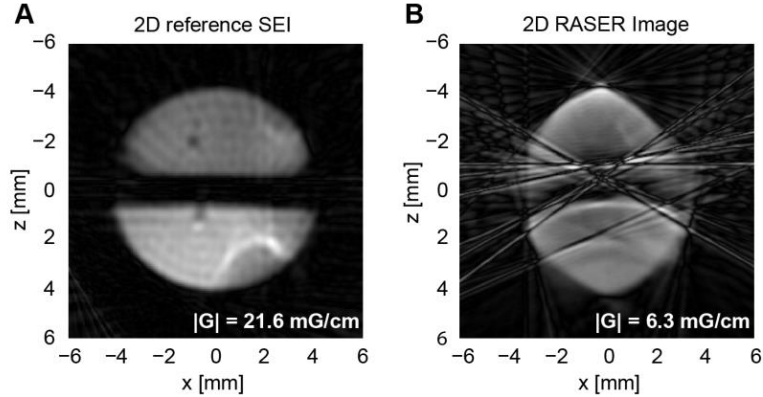

**Fig. 4. 2D SEI and 2D RASER image.** (**A**) 2D- SEI and (**B**) 2D RASER MRI measured at 3.9 mT. The 2D images (**A**) and (**B**) are obtained by projection reconstruction of 30 projections each. These 1D projections are measured with the sequence in Fig.2(C, D) from different angles by varying $G_x$ and $G_z$ such that $G_x^2+G_z^2$ = const. In (**A**), the two capillaries used for *p*-H$_2$ supply are visible around x = –1 mm, z = 0.5 mm and x = –1.5 mm, z = –2 mm for each chamber. The RASER image (**B**) is recorded at a 3.5 times smaller gradient than (**A**), but both spatial resolutions are similar. The RASER image is affected by interference lines. The origin of these artifacts is discussed in the text and in the SI, section 5.

A stark contrast of RASER MRI to traditional MRI is the dependence of RASER MRI images on the magnitude of the nuclear spin polarization. Figure 5 shows a series of 1D RASER and SEI images of the phantom, acquired with decreasing levels of polarization, i.e. decreasing population inversion $d_0$. The polarization was adjusted by implementing an increasing waiting time $\Delta t$ between the polarization step and acquisition.

For SEI, decreasing polarization entails decreasing SNR for each image in Fig. 5(A), but the shape of the image in the interval 2s < $\Delta t$ < 20 s (about a few $T_1$ relaxation periods) remains invariant. The spatial resolution for the SEI is determined by the slope on the sample boundaries with $\delta z_{SEI} \approx$ 50 μm. This observation is in overall good agreement with the theoretical expectation of $\delta z_{SEI}$ = w/($\gamma_H G_z$) = 55 μm. Although the initial negative polarization ($d_0$) changes by more than a factor 10 within the first 20 s, the shape of the SEI images is invariant. This behavior is because the widths of the underlying PSFs barely deviates from a Lorentzian line-width and radiation damping effects are insignificant. At longer waiting times ($\Delta t$ > 20 s), noise becomes more dominant and the shape deteriorates as more efficient relaxation at the walls decreases the image amplitude at the boundaries of the sample.

**RASER MRI dependence on polarization**

In contrast, the RASER image shape in Fig. 5(B) strongly depends on polarization. We attribute the differences between the two image halves to disparities in the bubbling rates and phantom shapes (details see SI section 4(c)). Fig. 5(C) shows simulated RASER images for five different initial population inversions $d_0$ and corresponding profiles $\rho_d(v)$ (see SI, Fig. S10) to examine the origin of the RASER image distortions. The experiment at $\Delta t = 8$ s matches the simulation with only one peak (width = 0.6 Hz, SI, Fig. S9) and for the experiments $\Delta t < 8$ s, the simulation qualitatively reflects the amplitude deformations and side lobes seen in the measured images. The ripples in some images in Fig. 5(B) cannot be simulated assuming a uniform division of the RASER image into $N = \Delta/\delta v$ slices. Motional artifacts and variations of $T_1$, $T_2^*$, and $B_1$-field over the image domain may be responsible for the observed ripples.

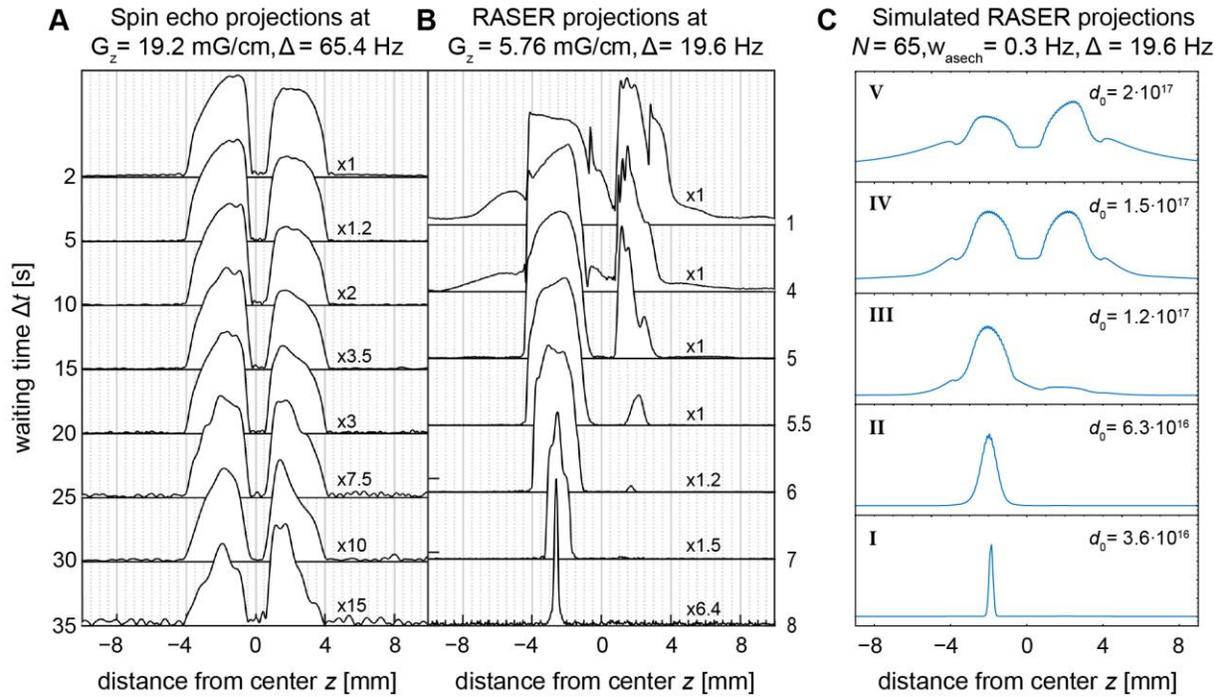

**Fig. 5. Projections of measured SEI, RASER MRI at $B_0 = 7.8$ mT and simulated RASER MRI at different waiting times $\Delta t$.** (**A**) The SEI was acquired at $G_z = 19.2$ mG/cm ($\delta z_{SEI} = 0.055$ mm) without slice selection. The shape remains form invariant up until $\Delta t = 20$ s. (**B**) The 1D RASER image was acquired at $G_z = 5.76$ mG/cm. At higher polarizations, i.e. for $\Delta t < 5$ s, both sides of the image are governed by strong nonlinear effects. At lower polarization, $\Delta t > 5$ s, the amplitude of right half in the phantom is strongly attenuated. At $\Delta t = 8$ s, the RASER image is reduced to one peak of 0.6 Hz width. (**C**) Simulated RASER images, based on Eqs.(1-4) and on a profile $\rho_d(v)$ similar to the SEI in Fig. 5(A). These reflect the basic features at different values $d_0$ (I-V), i.e. side lobes outside of the imaging domain and nonlinear deformations. All spectra are phased in absolute mode and normalized to the maxima of each image.

## Conclusion

The proof of principle experiments provided here and the corresponding nascent theoretical framework motivate several new challenges, and promise an opportunity to overcome current fundamental limitations in MRI, potentially leading towards MRI with better signal to noise ratio and spatial contrast. There is absolutely no background signal from other protons (*e.g.*, water) because there is no RF excitation and the RASER signals solely stems from negatively polarized molecules at low concentration (*45*). The results of this work pave the way to background-free clinical RASER MR imaging with new and more sensitive contrast mechanism. Moreover, the absence of RF excitation (and by extension virtually zero Specific Absorption Rate (SAR)) and the use of substantially lower field gradients requirements (and by extension significantly reduced concerns over peripheral nerve stimulation) in RASER MRI offers new unprecedented standards of patient safety (*46*).

Furthermore, RASER MRI theory is connected to many seemingly disjunct applications in science and technology. The developed system of differential equation Eqs.(1-4) and its solutions for the RASER MRI model are equivalent to the fundamental equations in many other fields (see SI section 6) with prominent examples of synergetics (*15*) and non-linear dynamics (*14, 34, 47, 48*).

## Acknowledgments:

SL greatly acknowledges the RWTH Aachen University for accepting me as a guest scientist, providing the research environment and equipment to run all experiments of this study at the ITMC (Institute of Technical and Macromolecular Chemistry). SA thanks his wife Jenny Oquendo Mora for keeping up the moral during difficult situations in the COVID 19 pandemic. Excellent cooperative and IT support from Stefan van Waasen, Michael Schiek and Ulrich Probst from Forschungszentrum Jülich is greatly acknowledged. Michael Adams is greatly acknowledged for valuable help in designing the phantom.

## Funding:

Department of Defense CDMRP W81XWH-20-10576 (EYC)
National Heart, Lung, and Blood Institute 1 R21 HL154032-01 (EYC)
National Science Foundation CHE-1904780 (EYC)
National Institute of Biomedical Imaging and Bioengineering 1R01EB029829 (EYC, TT)
Office of Biological and Environmental Research of the U.S. Department of Energy Atmospheric System Research Program Interagency Agreement grant DE-SC0000001
National Institute of Health R21-EB025313 and R01EB029829 (TT)
The content is solely the responsibility of the authors and does not necessarily represent the official views of the National Institutes of Health.
TT acknowledges funding from the Mallinckrodt Foundation.

**Author contributions:**

    Conceptualization: SL, SA
    Methodology: SL, SA
    Software: SL, LL, SF, SA
    Resources: AA, TT, SA
    Formal Analysis: SL, SF, LL, SA
    Investigation: SL, SF, LL, SA
    Visualization: SL, SF, LL, TT, SA
    Funding acquisition: AA, TT
    Project administration: SA
    Writing – original draft: SL, TT, SA
    Writing – review & editing: SF, LL, MSR, EYC, AA

**Competing interests:**

TT is a founder, equity holder, and President of Vizma Life Sciences LLC (VLS). VLS is developing products related to the research being reported. The terms of this arrangement have been reviewed and approved by NC State University in accordance with its policy on objectivity in research. MSR is a founder and equity holder of VLS. EYC declares a stake of ownership in XeUS Technologies LTD. All other authors have no competing interest.

**Data and materials availability:**

The data that support the plots within this paper and other findings of this study are available from the corresponding author upon reasonable request.

**Supplementary Materials:**

    Materials and Methods
    Supplementary Text
    Figs. S1 to S11

**Materials and Methods**

SABRE samples were prepared under Schlenk conditions. The samples contained 5 mmol/l [Ir(cod)(IMes)Cl] of the SABRE catalyst precursor (*41*), and $c_{pyr}$ =100 mmol/l of pyrazine in methanol-$d_4$. Pyrazine was chosen because it is associated with a single resonance in the NMR spectrum with $ns_{pyr} = 4$ chemically and magnetically equivalent protons, ideal for RASER and spin echo imaging experiments. For single chamber experiments, the sensitive volume was filled with 900 μl solution, while for the experiments using two chambers with each 300 μl were filled into each side giving a total sample volume $V_s = 0.6$ ml. A glass capillary (~100 μm outer, 30 μm inner diameter), was introduced into each chamber for parallel $p$-$H_2$ supply. During polarization build-up, $p$-$H_2$ was bubbled though the solution at ~ 30 sccm and at 2 bar pressure. Parahydrogen was generated using a Bruker $p$-$H_2$ generator at 35 K, yielding ~ 94% enriched $p$-$H_2$ gas. Typically a negative pyrazine proton polarization of $P_H \approx -10^{-3}$ to $-10^{-2}$ is achieved in a magnetic field ranging from 3.9 to 7.8 mT. This chosen magnetic fields are close to the field $B_0 = 6.5$ mT, where the SABRE $^1$H polarization for pyridine and similar chemical motives such as pyrazine is maximized (*18*). With respect to RASER MRI low magnetic fields do offer the additional advantage of lower susceptibility artefacts.

A SABRE induced $^1$H polarization of $P_H = -10^{-3}$ corresponds to a population inversion $d_0 = c_{pyr} \cdot V_s \cdot (-P_H) \cdot ns_{pyr} = 0.1$ mol/l $\cdot 6 \cdot 10^{-4}$ l $\cdot -(-10^{-3}) \cdot 4 = 1.4 \cdot 10^{17}$. The total number of $^1$H spins in the sample is $N_s = 1.4 \cdot 10^{20}$. Analogue calculations yield the initial conditions for simulations in the main text and the SI. For example in Fig. 5 of the main text, the initial population inversion is assumed between $d_0 = 3.6 \cdot 10^{16}$ and $2 \cdot 10^{17}$. The $^1$H NMR parameters of pyrazine were measured to be $T_2^* = 0.7$ s (Lorentzian width w $= 1/(\pi T_2^*) = 0.455$ Hz). $T_1$ values at different positions were measured using the results of the SEI images versus $\Delta t$ (see Fig. 5(A)). We found $T_1 = 5.0$ s in the bulk. The measurement close to the walls varied around $T_1 = 2.5 \pm 0.5$s. For the simulations we chose a difference in $T_1$ between the bulk and the walls of 3 s.

The sample is located in a cylindrical glass tube (ID = 8 mm), divided by a glass slide (1 mm thickness) for two-chamber experiments. The designed phantom is hand-made. The 1mm thick glass sheet is held in place by chemically resistant glue. The liquid sample inside the two chambers is located in the sensitive volume of a cylindrical NMR detection coil (ID = 10 mm, height = 10 mm) which is connected to an external resonator with high quality factor (EHQE, $Q_{ext} = 360$ @ 166 kHz) for sensitive detection of the NMR or RASER signals (*49*). The total quality-factor of the combined resonator (external resonator and NMR coil) is $Q = 100$. The $B_1$ field profile from the NMR detection coil in the center of the sample is calculated to be about 10% lower compared to the field at the edges of the sample. As the RASER active slices interact through the $B_1$ field of the coil, the coupling now depend on space, which is not accounted for in the parameter $\beta$ in Eqs.(1-3). In summary, the dependence of $B_1$, $T_2^*$ and $T_1$ on the location of the sample are major sources for RASER imaging artifacts. Correction algorithms for artifacts are state of the art for high field MRI scanners(*50*) and could mostly be adapted to the artifacts presented here. The magnetic fields of the low cost MRI system are generated by a set of four handmade shim gradients ($G_x$, $G_y$, $G_z$ and $G_{crush}$) and by an electromagnet producing a constant field of the range of 0.5 mT < $B_0$ < 20 mT. For our experiments we chose $B_0 = 3.9$ mT and 7.8 mT corresponding to 166.6 kHz and 333.3 kHz $^1$H resonance frequency. The rf frequency of the spectrometer is chosen such that the off-resonance frequency $v_0$ is between 20 Hz and 150 Hz away from the $^1$H resonance frequency. The homogeneity of the $B_0$ field is 1 ppm/cm$^3$. The $p$-H$_2$ supply in a low-field electromagnet in conjunction with sensitive EHQE detection avoids the necessity of a shuttling system for rapid transport of the sample into a high-field magnet. The $G_x$ and $G_z$ gradients were used to obtain projections from 30 different angles (in 6 degree steps). All data was acquired in a single scan. Spin echo images were acquired at an echo time of 1 s. 2D images were obtained after projection reconstruction of the 1D slices using a matlab code, written for this project. The spatial resolution is divided into a resolution along a slice in radial and angular direction. The radial resolution is 50 μm for SEI at 21.6 mG/cm, which corresponds to 160 points along the 8 mm sample diameter. The angular resolution with 30 slices spanning 180° is 6°.

There are frequency shifts due to slow magnetic $B_0$ field drifts in the order of a few ppm per minute. At 333 kHz (7.8 mT) these drifts on a time scale of 10 minutes were more pronounced compared to 166 kHz (3.9 mT). The reason is thermal instability of the current supply in conjunction with heating of the resistive $B_0$ field coil. For one 1D RASER image measured at 7.8 mT with a corresponding RASER burst lasting a few seconds a drift of a few ppm per minute means less than 0.1 ppm or 0.03 Hz frequency drift. The image domain $\Delta$ is typically chosen between 10-100 Hz (corresponding to about 20-200 slices for SEI), so the drift for a single 1D RASER image is negligible. For a 2D RASER image with a total measuring time of about 30 min for all 30 1D slices, the central frequency between the individual 1D slices could differ by a few Hz. Thus, each 1D image was shifted to yield the same center frequency for all 1D images before projection reconstruction.

The simulations based on the model Eqs.(1-4) were performed using Mathematica 8. The NDSolve[] routine was used for the numerical evaluation of the variables $d_\mu(t)$, $A_\mu(t)$, and $\phi_\mu(t)$. The computation time of the system Eqs.(S5-S8) can be quite long depending on the number of modes $N$. All parameters $d_\mu$, $A_\mu$ and $\phi_\mu$ are coupled in between each other in a nonlinear way by the cos- and sin-terms on the right sides of Eqs.(S5-S7). This is the reason for many non-linear phenomena which can arise in RASER MRI model, ranging from phase locking, collapse phenomena, non-linear image distortions, edge effects, multiple-period doubling and chaos. While there are exactly $N$ coupling terms for $A_\mu$ and $\phi_\mu$ in Eqs.(S6, S7), the number of coupling terms for $d_\mu$ in Eq.(S5) is $N(N-1)/2$. For larger number of slices, $N > 100$, the system of equations becomes elaborate and a large amount of computation is required. In fact the computation time is roughly proportional to $N^3$, so the system Eqs.(S5-S7) is classified as a polynomial problem. A typical numerical evaluation using an I5 processor with 8 GB RAM takes about 60 s for $N = 50$ and can be many hours to days for $N > 100$.

For these simulations, initial conditions for $d_\mu(0)$, $A_\mu(0)$, and $\phi_\mu(0)$ are required. The initial conditions for $d_\mu(0)$ at $t = 0$ were calculated for a given profile $\rho_d(v)$ (Eq.(4)). For $N_s = 1.4 \cdot 10^{20}$ $^1$H spins the average value for the initial spin noise amplitude is $<A> \sim (N_s)^{1/2} = 1.18 \; 10^{10}$ with a random phase $\phi_\mu(0)$. For the simulations, constant values were assumed for simplicity (i.e. $A_\mu(0) = 10^{12}$, and $\phi_\mu(0) = 0$) since the RASER image is independent on the initial transverse spin components (see invariance principle III, main text and SI section 1).

# Supplementary Materials for

RASER MRI: Magnetic Resonance Images formed Spontaneously exploiting Cooperative Nonlinear Interaction

Sören Lehmkuhl[1,2]*, Simon Fleischer[3], Lars Lohmann[3], Matthew S. Rosen[4,5], Eduard Y. Chekmenev[6,7], Alina Adams[3], Thomas Theis[1,8,9]*, Stephan Appelt[3,10]*

Correspondence to: lehmkuhl@kit.edu, ttheis@ncsu.edu, st.appelt@fz-juelich.de

**Supplementary Text**

This supplement discusses the theory and the physics of RASER modes in the presence of a magnetic field gradient. It aims to explain many phenomena for RASER MRI in one dimension. For this purpose the supplement is subdivided into six subsections. We first introduce a model to extend the nonlinear multi-mode RASER theory (*12, 16*), which describes $N$ nonlinear interacting RASER modes (*9, 15, 20*). Section 1 describes the derivation of the equations of motion governing RASER MRI in one dimension. A spatial encoding procedure is added which divides the imaging domain $\Delta = \gamma_H G_z L$ in $N$ nonlinear interacting slices, where $\gamma_H$, $G_z$ and L are the $^1$H gyromagnetic ratio, the magnetic field gradient $dB_0/dz$, and the sample extension, respectively. In section 2, we derive the point spread function (PSF) for one single RASER mode or slice ($N = 1$) excluding $T_1$ relaxation: The hyperbolic secant. In section 3, $T_1$ relaxation is included and it is shown that the corresponding PSF as a function of time is an asymmetric hyperbolic like secant. In section 4, RASER MRI simulations for $N > 20$ interacting slices are discussed, especially for a rectangular profile (4a), a rectangular profile superimposed by a sinusoidal modulation (4b) and for a profile close to our experimental two-chamber setup (4c). In section 5, possible artifacts from the 2D and 1D RASER images from the main text are discussed. Finally, section 6 concludes with a discussion of the relation between RASER MRI and other fields of science and mathematics.

### 1. Theory of one-dimensional RASER imaging with $N$ interacting modes

As a starting point, one dimensional imaging is discussed. In this case, only one magnetic field gradient $G_z = dB_0/dz$ is applied and the resulting images are one-dimensional projections. We assume that the sample has been polarized into a state of negative spin polarization. In this manuscript, the sample is pumped by SABRE, but nothing precludes the use of other hyperpolarization techniques. We additionally assume, that there is no additional pumping (the pumping period is over) and there was a small but sufficient waiting time to assure that there is no more movement in the sample (see Fig. 2, main text). This will be the initial time at $t = 0$ which is characterized by a total population inversion $d_0$. In a one-dimensional model, the sample is characterized by its extension $L$ along the z-direction, a center $z_0$ and two boundaries at positions $z_0 + L/2$ and $z_0 - L/2$, as shown in Fig. S1(A).

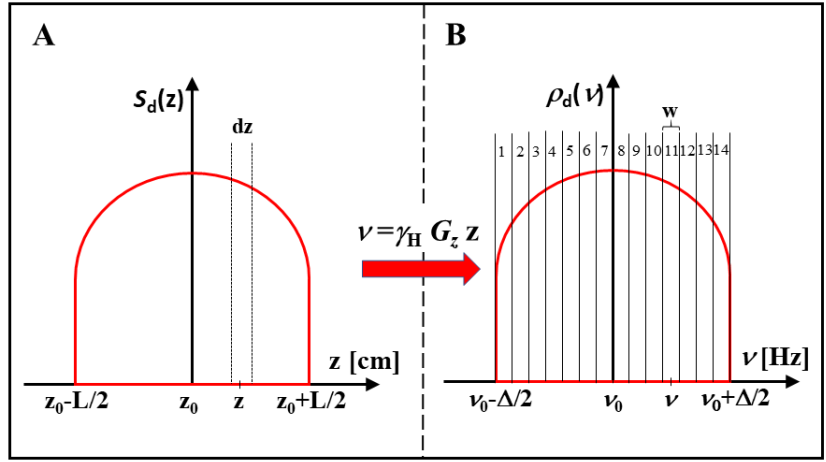

**Fig. S1. Transformation of the imaging profile from space into the frequency domain.** (**A**) Example of a one dimensional image with a circular shaped top, where $S_d(z)dz$, the number of inverted spins per slice thickness dz at position z, is plotted against the spatial z-coordinate. The total extension of the sample is L, $z_0$ denotes the center, the left and right boundaries are located at $z_0 - L/2$ and $z_0 + L/2$. (**B**) Given a gradient $G_z$ the density $S_d(z)$ transforms into the density $\rho_d(\nu)$ after a linear transformation $\nu = \gamma_H G_z z$ is performed onto the z-axis. $\rho_d(\nu)d\nu$ is defined as the number of inverted spins at frequency $\nu$ in the interval $d\nu$. The center frequency of the sample is $\nu_0 = \gamma_H B_0/2\pi$, the left and right boundaries are located $\nu_0 - \Delta/2$ and $\nu_0 + \Delta/2$, where the image domain is defined as $\Delta = \gamma_H G_z L$. Here the image is divided into $N = 14$ slices, where the width of each slice is given by the line-width $w = 1/(\pi T_2^*)$.

If the distribution of the population inversion at position z is the one-dimensional density function $S_d(z)$, then $S_d(z)dz$ is the number of negatively polarized spins in the spatial interval dz at the location z. In the presence of the gradient $G_z$ the z-dimension is transformed into the frequency space according to a linear transformation $v = \gamma_H \cdot G_z \cdot z$. After this linear transformation the density $S_d(z)$ transforms into the one dimensional density $\rho_d(v)$, where $\rho_d(v)dv$ is the number of negatively polarized spins at the frequency interval $[v,v+dv]$. The function $\rho_d(v)$ is shown on the right side of Fig. S1 with a center frequency $v_0$ and the two boundaries at $v_0+\Delta/2$ and $v_0-\Delta/2$. The image domain in frequency space is given by $\Delta = \gamma_H \, G_z \, L$. Given the two density functions $S_d(z)$ and $\rho_d(v)$ in the spatial and frequency domain the total initial population inversion $d_0$ of the sample is given by the integral

$$d_0 = \int_{z_0-L/2}^{z_0+L/2} S_d(z)dz = \int_{v_0-\Delta/2}^{v_0+\Delta/2} \rho_d(v)dv \quad (S1).$$

An assumption made for the simulations RASER MRI is that RASER activity in the presence of $G_z$ is self-organized. Thus, $N \gg 1$ modes or slices are introduced to enable numerical simulations of the theory. A good estimate for the number of slices is given by $N = \Delta/\delta v$, where the slice separation is chosen $\delta v < w$. We found by numerical evaluation that if $\delta v < w$, the properties and the shape of the resulting RASER images do not change (invariance principle I). For a simulation of a RASER image without numerical artifacts in the image domain $\Delta = \gamma_H \, G_z \, L$, a division into $N = \Delta/\delta v \sim \Delta/w_{as}$ slices is sufficient, where $w_{as} \sim 1/T_1$ is the width of the asymmetric point spread function of a single RASER mode (see section 3). Each RASER-active slice with label $\mu = 1..N$ can be attributed to a center frequency $v_\mu$ given by

$$v_\mu = v_0 - \frac{1}{2}\left[\Delta - (2\mu-1)\delta v\right], \quad \mu = 1...N \quad (S2).$$

The ensemble RASER image can be described by a superposition of $N = \Delta/\delta v$ slices. In contrast to regular MRI with its Lorentzian shaped PSFs, the spectrum associated to each slice is very complicate, because all slices are all coupled between each other. This is shown in section 4(a).

In the following we demonstrate the essential steps leading to the RASER MRI equations, which describe the dynamics of all $\mu = 1..N$ RASER slices. We start with the multi-mode RASER theory as described previously (*12,16*). These are given by a set of $2N$ non-linear coupled differential equations for $N$ modes. The dynamics of each mode or of the spin species with angular frequency $\omega_\mu$ is described by the population inversion $d_\mu$ and the complex valued transverse spin component $\alpha_\mu = A_\mu \exp(i\phi_\mu)$, where $A_\mu$ and $\phi_\mu$ are the amplitude and phase of each mode, respectively (*12*):

$$\dot{d}_\mu = \Gamma_\mu \left(d_{\mu,0} - d_\mu\right) - \frac{d_\mu}{T_1} - 2\beta \sum_{v,\sigma=1}^{N} \left\{\alpha_v \alpha_\sigma^* + \alpha_v^* \alpha_\sigma\right\} \quad (S3),$$

$$\dot{\alpha}_\mu = \left\{-\frac{1}{T_2^*} + i\omega_\mu\right\}\alpha_\mu + \beta d_\mu \sum_{\tau=1}^{N}\alpha_\tau \quad (S4).$$

The form of these equations, which is formulated in the rotating frame, are based on the multi-mode Laser equations, as published by H. Haken (*15*), and by applying the adiabatic elimination of fast variables (the slavery principle) to the fast decaying electromagnetic field modes of the LC resonator (*12*). For one single mode ($\mu = 1$) the form of Eqs.(S3,S4) is identical to the extended Bloch equations for the longitudinal and transverse magnetization $M_z$ and $M_T$, respectively used in many radiation damping studies (*4, 24, 29, 30, 36, 37, 40*). The magnetizations for one mode are related to the variables $d_1$ and $\alpha_1$ through $M_z = -(\hbar\gamma_X/2V_s)d_1$ and $M_T = -(\hbar\gamma_H/V_s)\alpha_1$, respectively.

The coupling parameter in Eqs.(S3,S4) is given by $\beta = \mu_0 \hbar \gamma_H^2 Q/(4V_s)$, where $\mu_0$, $\hbar$ and $\gamma_H$ denote the vacuum permeability, Planck's constant and the $^1H$ gyromagnetic ratio. $V_s$ is the sample volume and $Q$ the quality factor of the resonator, respectively. The coupling constant $\beta$ is related to the resonator damping rate $\kappa_m = \omega_0/Q$ and to the magnetic coupling constant $|g_m|^2 = \gamma_H^2 \mu_0 \hbar \omega_0/(4V_s)$ by $\beta = |g_m|^2/\kappa_m$. The factors $1/T_1$ and $1/T_2^*$ represent the longitudinal and effective transverse relaxations rates. Each population inversion $d_\mu$ of mode $\mu$ in Eq.(3) is pumped by the rate $\Gamma_\mu$ towards the equilibrium population inversion $d_{\mu,0}$, decays with the rate $1/T_1$ and is diminished by the last term on the right side of Eq.(S3), which depends on the sum over all quadratic terms $\alpha_\nu \alpha_\sigma^* + \alpha_\nu^* \alpha_\sigma$. The transverse modes $\alpha_\mu$ in Eq.(S4) decay with the rate $1/T_2^*$ and oscillate with the off resonance angular frequency $\omega_\mu$. For NMR spectroscopy the origin of these frequencies could be different chemical shifts or splittings due to $J$-coupling. The last term in Eq.(S4), proportional to the term $\beta d_\mu \Sigma \alpha_\tau$, is the source for the RASER emission of mode $\mu$ and is responsible for collective phenomena. The term $\beta d_\mu \alpha_\mu$ for $\tau = \mu$ is responsible for RASER emission. The product $\beta d_\mu = \mu_0 \hbar \gamma_H^2 Q d_\mu/(4V_s)$ can be described as radiation de-damping or damping terms of mode $\mu$, depending on whether the sign of $d_\mu$ is positive or negative, respectively. This is analogue to radiation damping described in the modified Bloch equations (*4, 24, 29, 30, 36, 37, 40*). The cross terms, $\beta d_\mu \alpha_\tau$ for $\tau \neq \mu$, are not included in the extended Bloch equations, but are essential to describe the dynamics of RASER MRI. Their sum determines the time evolution of each of the amplitudes $A_\mu$ and phases $\phi_\mu$. This has various consequences for RASER images, such as edge artefacts, and non-linear amplitude deformations, as will be shown later.

The RASER MRI equations for one $^1H$ spin species can be derived based on Eqs.(S3,S4): First, the spin system is pumped into a state with highly negative $^1H$ polarization $P_H$, i.e. into a state with a large total population inversion $d_0$ (see Eq.(S1)). For $^1H$ SABRE pumped organic molecules such as pyrazine or pyridine, the proton polarization is typically in the range from $P_H \sim -10^{-3}$ up to $-10^{-1}$, which for our experimental conditions corresponds to values $d_0 \sim 10^{16}$-$10^{18}$ (for calculation see Materials and Methods section). To prevent any RASER emission, a strong crusher gradient or a detuned resonance LC circuit ensure that $\alpha_\mu = 0$ during the pumping. Thus, only the pumping and $T_1$ relaxation terms on the right side of Eq.(S3) are relevant during the pumping period. Second, after the pumping and the strong crusher gradient is switched off (all $\Gamma_\mu = 0$) and provided the weak gradient $G_z$ for imaging is not too strong, some slices start to be RASER active and oscillate in the presence of $G_z$. The angular frequency $\omega_\mu = 2\pi \nu_\mu$ of each RASER active slice is given by Eq.(S2). After splitting Eq.(S4) into a real part describing the amplitudes $A_\mu$ and an imaginary part, which describes the phases $\phi_\mu$, a set of $3N$ nonlinear coupled differential equations for the variables $d_\mu$, $A_\mu$ and $\phi_\mu$ is obtained, i.e.

$$\dot{d}_\mu = -\frac{d_\mu}{T_1} - 4\beta \sum_{\sigma,\tau=1}^{N} A_\sigma A_\tau \cos(\phi_\sigma - \phi_\tau) \qquad (S5),$$

$$\dot{A}_\mu = -\frac{A_\mu}{T_2^*} + \beta d_\mu \sum_{\tau=1}^{N} A_\tau \cos(\phi_\tau - \phi_\mu) \qquad (S6),$$

$$\dot{\phi}_\mu = 2\pi\{\nu_0 - 0.5[\Delta - \delta\nu(2\mu-1)]\} + \beta \frac{d_\mu}{A_\mu} \sum_{\tau=1}^{N} A_\tau \sin(\phi_\tau - \phi_\mu) \qquad (S7),$$

$$d_\mu(0) = \int_{\nu_0 - \Delta/2 + (\mu-1)\delta\nu}^{\nu_0 - \Delta/2 + \mu\delta\nu} \rho_d(\nu) d\nu \qquad (S8).$$

Eqs.(S5-S8) describe the model for RASER MRI. Eq,(S8) is a boundary condition which defines the initial population inversion $d_\mu(0)$. The presence of the weak gradient $G_z$ spreads the total population inversion $d_0$ into $N = \Delta/\delta\nu$ slices, where the initial value of $d_\mu(0)$ for each slice is given by Eq.(S8), namely the integral of $\rho_d(\nu)$ over the frequency range of slice $\mu$ with boundaries [$\nu_0 - \Delta/2 + (\mu+1)\delta\nu$, $\nu_0 - \Delta/2 + \mu\delta\nu$].

The density $\rho_d(v)$ can be seen as a given input profile after pumping of the sample, which depends on the shape of the imaged object. Given $\rho_d(v)$, the quantities $d_\mu$, $A_\mu$ and $\phi_\mu$ can be evaluated by numerical evaluation of Eqs.(S5-S7) with the given boundary conditions Eq.(S8), and the measurable total transverse RASER signal results as a superposition of all functions $\alpha_\mu = A_\mu \exp(i\phi_\mu)$.

$$Sig(t) = \sum_{\mu=1}^{N} A_\mu(t) \text{Re}\{\exp[i\phi_\mu(t)]\} \qquad (S9).$$

The Fourier transformation of the total signal $Sig(t)$ results in the RASER image. Note that no boundary conditions for the initial amplitudes $A_\mu(0)$ and phases $\phi_\mu(0)$ are given in Eq.(S8). The reason is that the resulting RASER images are practically independent to the initial values of $A_\mu(0)$ and $\phi_\mu(0)$, no matter whether $A_\mu(0)$ and $\phi_\mu(0)$ are random or defined values. Provided the condition $T_1 \gg T_2^*$ holds (which is mostly the case), extensive numerical simulations reveal three invariance principles with respect to the Fourier transformed RASER images in the absolute mode:
I. The RASER image does not depend on to the slicing $\delta v$ as long as $\delta v < w$. II. The RASER image contrast and resolution is independent of $T_1$, and III. The RASER image is invariant with respect to the initial conditions $A_\mu(0)$ and $\phi_\mu(0)$. These three invariance principles have significant consequences for RASER MRI.
The invariance from the choice of the slicing distance $\delta v$ (principle I) means that a continuous limit $N \to \infty$ exists, where the discrete variables $d_\mu(t)$, $A_\mu(t)$ and $\phi_\mu(t)$ become continuous variables. In this limit all sums in Eqs.(S5-S8) become integrals. Another consequence is that numerical simulations produce reliable results without changing the involved physics as long as $\delta v < w$. We found that ripple like artifacts arise in the image if $\delta v \sim w$ but the envelope of the image is the same compared to $\delta v \ll w$. However, if $\delta v \ll w$, the numerical simulations can become very time consuming. A good compromise is $\delta v \approx 1/T_1 \approx w_{as}$, where no ripples are visible.
The invariance of the RASER image shape and contrast on the value of $T_1$, (principle II), means that the contrast of RASER MRI cannot be associated to the width of single PSFs. This includes the width of the asymmetric PSF $w_{as} \approx 1/T_1$ of a single RASER mode, introduced in section 3. The contrast mechanism is based on collective interactions, as will be shown in chapter 4(a).
The invariance of the RASER image on the initial conditions $A_\mu(0)$ and $\phi_\mu(0)$ (principle III) allows for reproducible results irrespective of the noise excitation. The amplitude and shape of the RASER image do not change if the RASER burst is initiated either by spin noise or a defined excitation sequence with a weak RF or DC pulse. We found that for very different initial conditions $A_\mu(0)$, $\phi_\mu(0)$ (more than one order of magnitude), there is a small shift of the entire RASER burst signal. This leaves the absolute Fourier transformed spectrum invariant but the phased spectrum is shifted by a global phase. This small shift can be avoided by applying a weak DC or rf pulse to initiate the reproducible RASER bursts, beneficial for averaging and 2D RASER MRI.

A detailed mathematical analysis of the dynamics of the image formation and how to explain the three invariance principles is quite elaborate and not the main focus of this contribution. Therefore, we focus on numerical simulations to investigate image artifacts and the more sensitive contrast mechanism. In section 4(a) we use the simple example of a rectangular profile. Due to the invariance principle III, we chose the simplest case, $\phi_\mu(0) = 0$ and a constant small value $A_\mu(0) \sim 10^9$–$10^{10}$. The value for $A_\mu(0)$ is in the order of the spin noise amplitude $N_s^{1/2}$ (26), where $N_s \sim 10^{19}$-$10^{20}$ is the total number of spins in the sample.

Finally, we analyze whether a local threshold condition exists for RASER MRI. Without gradient, there is only one RASER mode ($N = 1$). The threshold condition for RASER action is $d_0 \geq d_{th} = 4V_s/(\mu_0 \hbar \gamma_H^2 T_2^* Q)$,(12) where $d_{th}$ is the threshold population inversion. An equivalent expression for this threshold condition is $\varepsilon = d_0/d_{th} = T_2^*/\tau_{rd} \geq 1$, where $\varepsilon$ is a dimensionless quantity characterizing the threshold. However if a gradient is applied, the sample divides into $N = \Delta/\delta v \gg 1$ slices and the threshold condition $\varepsilon = d_0/d_{th} > 1$ cannot be used any more as a proper criterion for RASER activity.

When a gradient is applied, the population inversion $d_0$ is distributed over the image domain $\Delta$ according to the population inversion density $\rho_d(v)$. Assuming no interaction between the slices $\delta v$, a threshold condition can be formulated based on the population inversion within a frequency interval $w = 1/(\pi T_2^*)$. Within this frequency interval the damping with rate $1/T_2^*$ is compared to the RASER dedamping process. A region $r_\mu = [v_\mu-w/2, v_\mu+w/2]$ at position $v_\mu$ and width $w = 1/(\pi T_2^*)$ is RASER active if $\int_{v_\mu-w/2}^{v_\mu+w/2} \rho_d(v) dv \geq d_{th}$.

One simple example is the case of a rectangular profile, where $\rho_d(v) = \rho_d^{rect} = d_0/\Delta$. Now, the integral becomes $\int_{v_\mu-w/2}^{v_\mu+w/2} \rho_d(v) dv = \rho_d^{rect} \cdot w = d_0 w/\Delta$, and the above threshold condition reduces to $d_0 w/\Delta \geq d_{th}$. This expression can be written as $\rho_d^{rect} \geq \rho_d^{th}$, where $\rho_d^{th} = d_{th}/w$ is the threshold population inversion density. Consequently if $\rho_d^{rect} < \rho_d^{th}$, no RASER action should be possible for any region $r_\mu$.

Unfortunately, this condition is not sufficient as a strict threshold condition in the presence of nonlinear coupling. Numerical simulations show that if $\rho_d^{rect}$ is slightly below $\rho_d^{th}$ a region $r_\mu$ in the center of $\Delta$ can be RASER active while the regions at the boundaries are not. Close to the center the slices cooperate with all its neighbors in a constructive way, while the slices close to the two boundaries cooperate destructively with their neighbors. Because of the cooperative action between slice $\mu$ with all other slices, a local threshold condition does not exist and has to be replaced by a non-local threshold condition. Numerical simulations show that RASER action for a region $r_\mu$ in the domain $\Delta$ depends on $d_0$, on the width of the image domain $\Delta$ and on the detailed shape of $\rho_d(v)$. A detailed mathematical evaluation of this statement is quite elaborate, but a numerical evaluation of the case using a rectangular profile is shown in section 4(a). In nearly all simulations shown in the following the threshold population inversion density $\rho_d^{th} = 4\pi V_s/(\mu_0 \hbar \gamma_H^2 Q)$ is used as a reference for RASER activity (even if not strictly valid) and indicated by a dashed green line.

A maximum gradient $G_{max}$ for RASER activity can be deduced from the threshold condition $\rho_d^{rect} = d_0/\Delta = \rho_d^{th}$ and in the absence of nonlinear coupling. From $\Delta = \gamma_H G_{max} L$ and $\rho_d^{th} = 4\pi V_s/(\mu_0 \hbar \gamma_H^2 Q)$ we obtain the maximum gradient $G_{max} = \mu_0 \hbar \gamma_H d_0 Q/(4\pi V_s L)$. This means that $G_{max}$ is large for high $Q$ resonators, large values of $d_0$ and if the sample is small.

## 2. RASER with $N = 1$, Point Spread Function neglecting $T_1$ relaxation

We will proceed with the simplest possible case, and derive the exact solution for one RASER mode ($N = 1$) if $T_1$ relaxation can be completely neglected ($T_1 = \infty$). In this case an exact solution in the form of tanh- and sech-functions exists. The relevance for the line shape in NMR spectroscopy due to strong radiation damping effects has been demonstrated (4, 24, 29, 30, 35-37, 40). Here, we repeat the derivation of the basic results described there for two reasons. The first reason is to be compatible with the nomenclature used here. Secondly, Mao *et al.* (29, 35, 36) studied amongst others the radiation damped signal burst after a radio frequency pulse excitation angle close to $\pi$, which is in close correspondence to one self induced RASER burst in Eqs.(S5-S8). For $N = 1$ and neglecting $T_1$ relaxation Eqs.(S5-S8) are reduced to its most basic form,

$$\dot{d} = -4\beta A^2, \qquad (S10),$$

$$\dot{A} = \left(\beta d - \frac{1}{T_2^*}\right) A, \qquad (S11),$$

$$\dot{\phi} = \omega_1, \qquad (S12).$$

We choose as boundary conditions for Eqs.(S10-S12) $\phi(0) = 0$, $A(0) = A_0 = 10^{12} \ll d_0$, $d(0) = d_0$. The two Eqs.(S11,S12) are equivalent to one single differential equation for the complex valued transverse spin component $\alpha(t) = A(t)\exp(i\omega_1 t)$, given by $\dot{\alpha} = (\beta d - 1/T_2^* + i\omega_1)\alpha$. The derivation of the exact solution for $d(t)$ and $A(t)$ follows from differentiation of Eq.(S10), which results in $\partial^2 d/\partial t^2 = -8\beta A \cdot \partial A/\partial t$. After substituting the right side of Eq.(S11) for $\partial A/\partial t$ in the preceding expression we obtain

$$\frac{\partial^2 d}{\partial t^2} = -8\beta A \cdot \left[\beta A \cdot d - \frac{A}{T_2^*}\right] \underset{\text{Eq.(S10)}}{=} \left(2\beta d - \frac{2}{T_2^*}\right) \cdot \frac{\partial d}{\partial t} \quad (S13).$$

Setting $U = \partial d/\partial t$, Eq.(S13) turns into $\partial U/\partial t = (2\beta d - 2/T_2^*)\partial d/\partial t$, which is equivalent to $\partial U/\partial d = 2\beta d - 2/T_2^*$. The solution is given by integration, which results in $U = \partial d/\partial t = \beta d^2 - (2/T_2^*)d + C$. The Integration constant $C$ is fixed by the initial conditions $A(0) \sim 0 \ll d_0$, $d(0) = d_0$, so $dd/dt = -4\beta A^2(0) \sim 0$, giving $C = d_0(2/T_2^* - \beta d_0)$. After introducing the parameter $q = (1 + \varepsilon^2 - 2\varepsilon)^{1/2}$, where the dimensionless variable is $\varepsilon = T_2^*/\tau_{rd} = T_2^* \beta d_0$, the previous equation for $U$ can be written as

$$U = \frac{\partial d}{\partial t} = \beta \left[-\left(\frac{q}{\beta T_2^*}\right)^2 + \left(d - \frac{1}{\beta T_2^*}\right)^2\right] \quad (S14).$$

Introducing the parameters $a = 1/(T_2^*\beta)$ and $b = q/(T_2^*\beta)$, Eq.(S14) simplifies to $\partial d/\partial t = \beta\left[-b^2 + (d-a)^2\right]$, which is equivalent to

$$\frac{1}{\beta}\int \frac{\partial d}{(d-a)^2 - b^2} = \int \partial t = (t - t_0) \quad (S15).$$

This integral is known as $\int dx/[(a-x)^2 - b^2] = b^{-1} \operatorname{arctanh}[(a-x)/b]$, so Eq.(S15) can be written as $\operatorname{arctanh}[1(1-\beta T_2^* d)/q] = (t-t_0)q/T_2^*$. After isolating $d$ on the left side we get

$$d = \frac{1}{\beta T_2^*}\left\{1 - q\, \tanh\left[\frac{q}{T_2^*}(t-t_0)\right]\right\} \quad \text{(S16)}$$

The exact form for the transverse spin component $A(t)$ can be derived by differentiation of Eq.(S16). Using the relation $\partial \tanh(x)/\partial x = 1 - \tanh^2(x) = \operatorname{sech}^2(x)$ this leads to

$$\frac{\partial d}{\partial t} = \frac{-1}{\beta T_2^*}\frac{q}{T_2^*} q\left\{1 - \tanh^2\left[\frac{q}{T_2^*}(t-t_0)\right]\right\}. \quad \text{(S17)}$$

According to Eq.(S10) $\partial d/\partial t = -4\beta A^2$ so Eq.(S17) can also be written as $A^2 = \beta^{-2}(q/2T_2^*)^2 \operatorname{sech}^2[q(t-t_0)/T_2^*]$. Applying the square root on both sides of the preceding equation, $A$ is finally

$$A = \frac{q}{2\beta T_2^*}\operatorname{sech}\left[\frac{q}{T_2^*}(t-t_0)\right] \quad \text{(S18)}.$$

According to Eq.(S18), $A$ is a hyperbolic secant function (soliton solution in the time domain) and is symmetric with respect to $t_0$, which is the time of maximum amplitude. An interesting feature of Eq.(S18) is that the envelope of the Fourier transformed spectrum in the frequency domain $\omega$ is once again a sech-function. According to Mao et al. (35, 36) this envelope is modulated by a phase factor $\cos(\omega t_0)$, i.e.

$$S(\omega) = \frac{\pi}{2\beta}\frac{q}{T_2^*}\operatorname{sech}\left[\frac{\pi T_2^* \omega}{2q}\right]\cos[\omega t_0] \quad \text{(S19)}.$$

The Fourier transformed spectrum of the transverse spin component $\alpha = A\exp(i\omega_1 t)$ is obtained from Eq.(S19) simply by replacing the angular frequency $\omega$ in the argument of the sech function and in the cos term by $\omega - \omega_1$ (Fourier shift theorem). By inspection of Mao et al. (35, 36) it can be shown that for the case of a self-induced RASER burst, the time $t_0$ is given by

$$t_0 = -\frac{T_2^*}{q}\operatorname{arctanh}\left[\frac{1+\varepsilon\cos\theta_0}{q}\right] \quad \text{(S20)}.$$

The exact expression for the factor is $q = \sqrt{1+\varepsilon^2 + 2\varepsilon\cos\theta_0}$, $\varepsilon = T_2^*/\tau_{rd} = T_2^*\beta d_0$ and the initial flip angle is given by $\theta_0 = \pi - 2\arcsin(A(0)\beta T_2^*/\varepsilon)$. The initial flip angle $\theta_0$ after applying rf-pulses close to 180°, as discussed in Mao et al., is replaced for the RASER by a small initial fluctuation of the transverse spin component $A(0) = A_0$ at time $t = 0$. This fluctuation initiates a self-induced single mode RASER burst in the absence of any rf-pulse. The sech-function Eq.(S18) is symmetric with respect to the time $t$, so we call this the symmetric Point Spread Function (PSF) which is valid for RASER imaging only if $T_1 = \infty$. An important feature of the spectrum described by Eq.(S19) is that close to the RASER threshold, i.e. $\varepsilon =$

$T_2^*/\tau_{rd} = T_2^* \beta d_0 \approx 1$ the factor $q \ll 1$ becomes very small, so the associated linewidth $w_{sech}$ of the corresponding spectrum Eq.(S22) is much smaller compared to the linewidth $w = 1/(\pi T_2^*)$ of a Lorentz-shaped peak, the latter representing the standard PSF for Spin Echo Imaging (SEI). The full width at half maximum $w_{sech}$ is determined by the argument $[\pi T_2^* \omega /2q]$ in Eq.(S19), which for $\varepsilon \approx 1$, $\theta_0 \approx \pi$ and $w = 1/(\pi T_2^*)$ is $w_{sech} = (2/\pi) \ln(2+3^{1/2})(\varepsilon - 1) w = 0.84(\varepsilon - 1)w$. For $\varepsilon > 2.2$ the width $w_{sech} > w$, while in the range $1 < \varepsilon < 2.2$ closer to threshold $w_{sech} < w$. For example, at $\varepsilon = 1.4$ $w_{sech} = 0.336$ $w = 0.15$ Hz for $T_2^* = 0.7$ s. The argument $w_{sech} < w$ for $1 < \varepsilon < 2.2$ holds even for the case of finite $T_1$ relaxation associated with an asymmetric PSF, as will be shown in the next section.

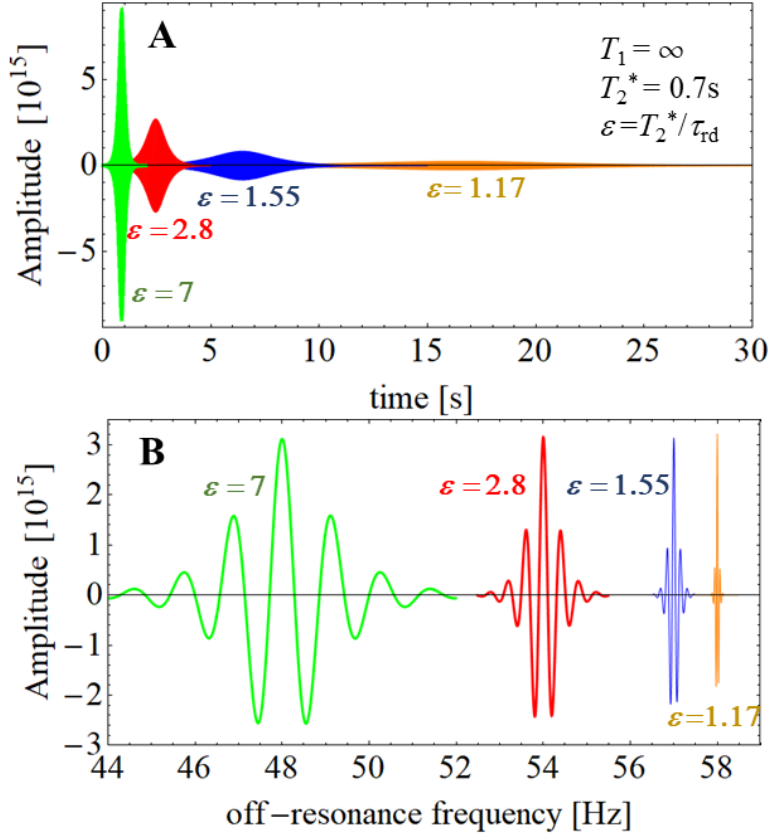

**Fig. S2. Numerical simulation of RASER bursts for a single mode ($N = 1$).** Simulation parameters: $T_1 = \infty$, $T_2^* = 0.7$ s, $Q = 100$, $V_s = 0.5$ cm$^3$. Initial values: $\phi(0) = 0$, $A(0) = 10^{13}$, and four different values for $\varepsilon = T_2^*/\tau_{rd} = T_2^* \beta d_0$ are assumed. (A) RASER burst versus time for $\varepsilon = 7$ (green), 2.8 (red), 1.55 (blue) and 1.16 (orange). All four simulated signals are symmetrical sech-functions, which are identical to the analytical hyperbolical secant solution given by Eq.(S21) and which are symmetric with respect to the time $t_0$ (the time where the maximum signal appears). (B) Corresponding simulated phased Fourier-transformed spectra of the four signals shown in Fig. S2(A). The four point spread functions (PSF`s) are displaced in the frequency dimension (with center frequencies at 48, 54, 57 and 58 Hz) for clarity. Note that in each PSF and with decreasing $\varepsilon$ the spectral width and the spacing between subsequent oscillations is decreasing.

Fig. S2(A) shows the numerical simulations of four different PSF for the case $T_1 = \infty$. The transverse spin component $A(t)$ versus time $t$ is plotted for four different parameters $\varepsilon = T_2^*/\tau_{rd} = T_2^* \beta d_0$, ranging from $\varepsilon = 7$ (green) far above threshold, $\varepsilon = 2.8$ (red), $\varepsilon = 1.55$ (blue) to $\varepsilon = 1.16$ (orange) close to threshold. The simulation parameters are $T_2^* = 0.7$s, $Q = 100$, $V_s = 0.5$ cm$^3$ and the initial conditions are $\phi(0) = 0$, $A(0) = 10^{13}$. All four simulated RASER bursts are symmetric sech-functions which are identical to the analytical solution given by Eq.(S19). Furthermore, each PSF is symmetric with respect to the time of maximum signal, $t_0$. The time $t_0$, as given by Eq.(S20), and the width of each burst increases for decreasing values $\varepsilon$. As $\varepsilon$ approaches the threshold ($\varepsilon = 1$) the duration or width in the time domain of each PSF becomes arbitrarily long, and the maximum amplitude arbitrarily small, but the area below each of the RASER envelopes remains the same. The corresponding Fourier- transformed spectra are shown in Fig. S2(B). The four PSF`s are displaced in the frequency space for a better separation. Note that with decreasing $\varepsilon$ the spectral width and the period due to the modulation proportional to the factor $\cos(\omega t_0)$ (see Eq.(S19)) in each PSF is decreasing. Close to threshold the corresponding PSF are much narrower compared to the linewidth of a conventional Lorentz-peak. In our experiments typical measured values are $T_1 \sim 3$ s -5 s and $T_2^* = 0.7$ s, so the PSF has a width in the range $0.2$ Hz $< w_{as} < 0.33$ Hz and the Lorentzian line width is $w = 1/(\pi T_2^*) = 0.455$ Hz.

### Case $N = 1$ including $T_1$ relaxation: The asymmetric Point Spread (a-PSF).

For this next case, we include $T_1$ relaxation. To our knowledge this has not been discussed in the literature. The dissipation caused by $T_1$ relaxation, represented by the additional term $-d/T_1$, is introduced in Eq.(S10) for one mode ($N = 1$). No analytical solution exists for this case, and the symmetry of the PSF given by Eq.(S18) with respect to time $t_0$ is lost. Fortunately the key properties of the corresponding asymmetric PSF can be evaluated by numerical simulations of Eqs.(S10-S12), including the additional loss term $-d/T_1$.

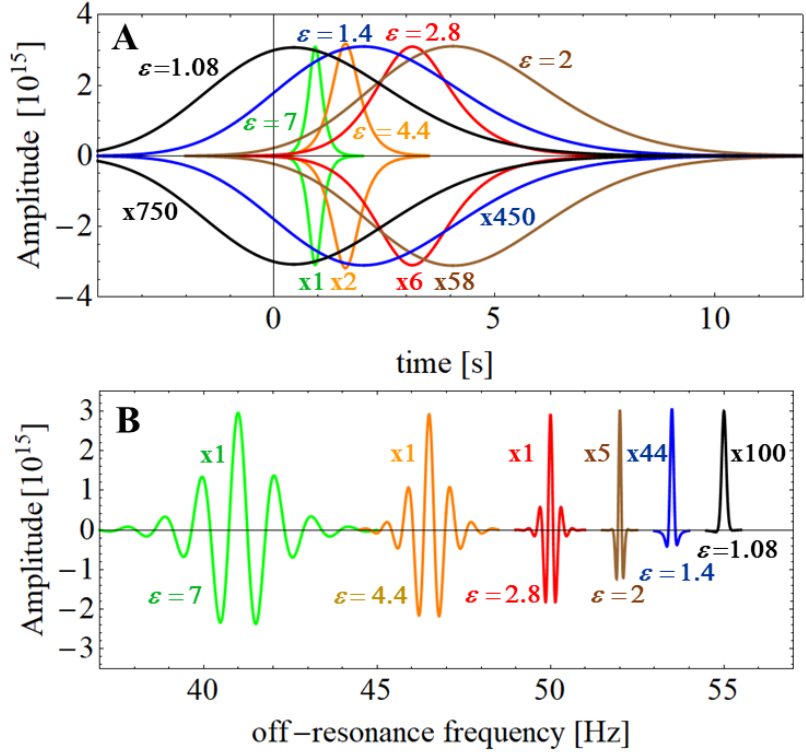

**Fig. S3.** Series of six simulated asymmetric Point Spread Functions (a-PSF) for $T_1 = 6$ s. (**A**) Plot of the envelopes of transverse spin components $A$ (RASER bursts) versus time for six different values $\varepsilon$. Simulation parameters: $T_1 = 6$ s, $T_2^* = 0.7$ s, $Q = 100$, $V_s = 0.5$ cm$^3$. Initial values: $\phi(0) = 0$, $A(0) = 10^{13}$, and $\varepsilon = T_2^* \beta d_0 = 7$ (green), 4.4 (orange), 2.8 (red), 2 (brown), 1.4 (blue) and 1.08 (black). Note the large differences in amplitude and duration of the corresponding a-PSF. (**B**) Corresponding phased Fourier-transformed spectra of the six a-PSF`s from Fig. S3(a). Far above threshold, at $\varepsilon = 7$, the peak amplitude including the modulation due to the $\cos(\omega t_0)$-term of the spectrum looks quite similar to the spectrum for $T_1 = \infty$ in Fig. S2(b). Close to threshold, at $\varepsilon = 1.4$ and 1.08, the peak amplitude is tens of times smaller compared to $\varepsilon = 7$ and there is nearly no phase modulation visible.

To find out how the a-PSF depends on the parameter $\varepsilon$ we performed various simulations as shown in Fig. S3, where a series of six PSF in the time and frequency domain are shown for different values of $\varepsilon$. In Fig. S3(A) the envelope of the amplitude of $A$ is plotted versus time for $\varepsilon =1.08$ (black), 1.4 (blue), 2 (brown), 2.8 (red), 4.4 (orange) and 7 (green). The simulation parameters are given in the caption of Fig. S3. Due to $T_1$ relaxation, all RASER bursts look like slightly asymmetric sech-functions (a-PSF). The weak asymmetry is with respect to their maximum signal at time $t_0$, i.e. the duration of the initial rise of all signals is slightly shorter compared to the decaying tail. $T_1$ relaxation is also responsible for large differences in the maximum amplitudes between each a-PSF. Close to threshold at $\varepsilon =1.08$, the maximum amplitude is several hundred times smaller in comparison to $\varepsilon = 7$, far above threshold. Conversely, the duration of the a-PSF at $\varepsilon =1.08$ is tens of times longer when compared to the duration at $\varepsilon = 7$. Fig. S3(B) shows a plot of the corresponding phased Fourier transformed spectra of the six a-PSF in Fig. S3(A).

Far away from threshold, at $\varepsilon = 7$, the peak amplitude and the modulation due to the $\cos(\omega t_0)$-term of Eq.(S22) of the spectrum looks quite similar to the spectrum for $T_1 = \infty$ in Fig. S2(B). At $\varepsilon = 1.4$ and 1.08 close to threshold, the peak amplitude is tens of times smaller compared to the peak amplitude at $\varepsilon = 7$. Additionally, close to threshold there is nearly no phase modulation visible in the spectrum, in contrast to the phased spectra close to threshold in Fig. S2(B), which are characterized by a narrow envelope modulated by several oscillations. This difference in the number of periods visible in the PSF in Fig. S2(B) and a-PSF in Fig. S3(B) close to threshold is directly connected to the time of maximum amplitude, $t_0$. To analyze this key property, the time $t_0$ as a function of $\varepsilon$ has been numerically evaluated for different values of $T_1$. The result is shown in Fig. S4, where the simulation parameters are given in the caption.

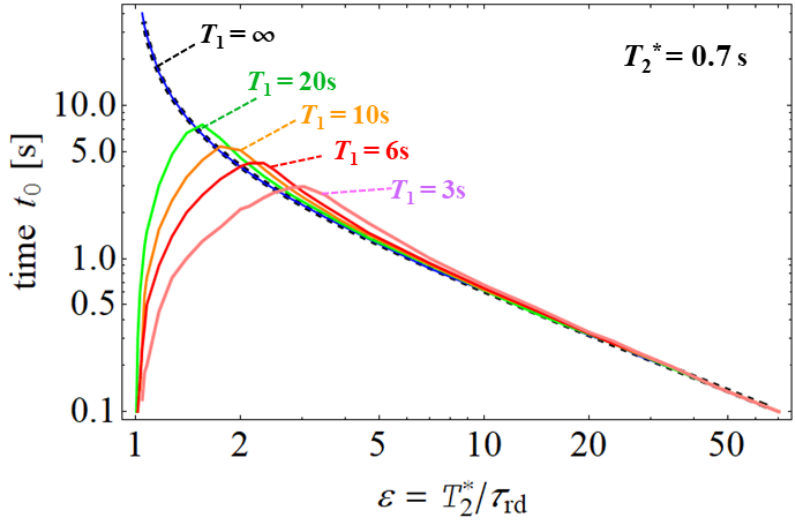

**Fig. S4. Time $t_0$ versus $\varepsilon = T_2^*/\tau_{rd}$ for five different values of $T_1$**. The dotted black and solid blue line correspond to $T_1 = \infty$ and represent the exact expression for $t_0$ (Eq.(S23)) and the $t_0$-value evaluated from numerical simulation, respectively. The four other plots represent $t_0$ derived from numerical simulations for $T_1 = 20$ s (green), 10s (orange), 6 s (red) and 3 s (pink). Simulation parameters are: $T_2^* = 0.7$ s, $Q = 100$, $V_s = 0.5$ cm$^3$. Initial values: $\phi(0) = 0$, $A(0) = 10^{13}$. For finite $T_1$ values there is a maximum in the evaluated function $t_0(\varepsilon)$ in the range $1.5 < \varepsilon < 4$. For larger $\varepsilon > 5$ further away from threshold, all curves asymptotically approach the curve for $T_1 = \infty$.

For the case $T_1 = \infty$, $t_0$ decreases monotonically with increasing $\varepsilon$. Note the dotted curve, which represents the exact solution given by Eq.(S20), is in good agreement with the numerical solution given by the solid blue line. A different behavior is observed for finite values of $T_1$, here ranging from $T_1 = 20$ s, 10 s, 6 s to 3 s. Starting at $\varepsilon = 1$, $t_0$ increases with increasing $\varepsilon$ until a maximum value is reached in the range $2 < \varepsilon < 4$ and finally $t_0(\varepsilon)$ decreases until approaching the curve for $T_1 = \infty$ in an asymptotic way. Here we state the similarity between the PSF and the a-PSF far from threshold and significant differences close to threshold.

# 3. Simulation of 1D RASER images for three spin-density profiles

Until now, we have studied PSFs for one single slice ($N = 1$). To describe RASER MRI, the concept of a single PSF is insufficient. Thus, in the following subsections, we will explore the physics of 1D RASER images consisting of many interacting slices ($N \gg 1$). Now collective effects dominate the physics of image formation and are essential to describe the image contrast as well as the nonlinear artifacts that arise. To understand these phenomena, we chose simple profiles to reproduce typical RASER MRI features in simulations (see subsections 4 (a-c)).

In all the simulations shown here, we assume parameters matching our experimental conditions: $T_1 = 5$ s, $T_2^* = 0.7$ s (line width w $= 1/(\pi T_2^*) = 0.455$ Hz), quality factor $Q = 100$, a cylindrical sample with volume $V_s = 0.5$ cm$^3$ and diameter $L = 0.8$ cm. All these parameters have either been measured in our $^1$H RASER-imaging experiments at 166 or at 333 kHz $^1$H Larmor frequency ($B_0 = 3.9$ or 7.8 mT) with SABRE pumped pyrazine, or correspond to parameters of the setup. The typical range for the magnetic field gradient is $2 \cdot 10^{-4}$ G/cm $< G_z < 2 \cdot 10^{-2}$ G/cm, which for a sample dimension of L = 0.8 cm corresponds to image domains in frequency space ranging from 0.67 Hz $< \Delta <$ 67 Hz. Three different spin density profiles are defined in Fig. S5: A rectangular profile above threshold, a profile with a constant value superimposed by a sinusoidal profile slightly below threshold, and a profile which is close to the experimental conditions with two sections with flat tops separated by a gap and with edges described by tanh functions.

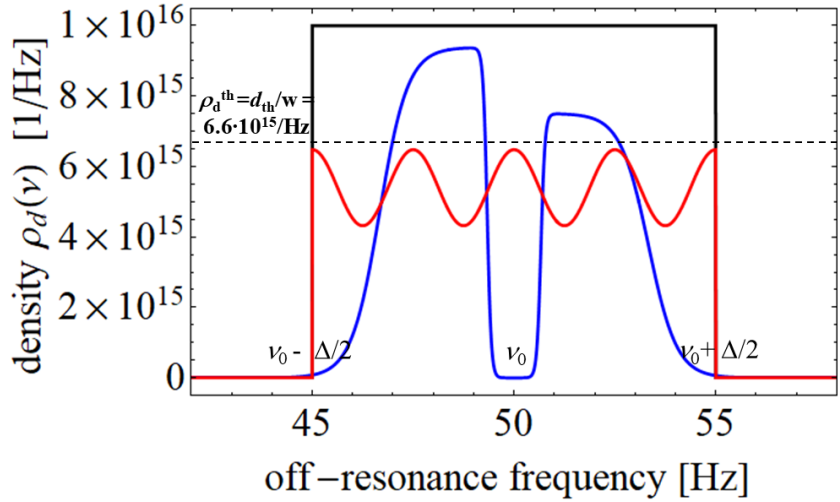

**Fig. S5. Three different density profiles $\rho_d(v)$ as possible inputs for RASER MRI**. All three profiles have an image domain of $\Delta = 10$ Hz centered at the central offset frequency $v_0 = 50$ Hz. In the interval 45 Hz $< v <$ 55 Hz the three profiles are: A rectangular profile $\rho_d(v) = \rho_d^{rect} = 10^{16}$/Hz, (black), a cosine-modulated profile with offset, $\rho_d(v) = 5.5 \cdot 10^{15}$/Hz $(1+ 0.2 \cos[8\pi(v-v_0)/\Delta])$ (red) and two sections with tanh shaped edges separated by a gap, $\rho_d(v) = 4.5 \cdot 10^{15}$/Hz $(0.5(\tanh[(v-v_0+\Delta/3)/0.7] - \tanh[(v-v_0+\Delta/15)/0.1]) + 0.4 (\tanh[(v-v_0-\Delta/15)/0.1] - \tanh[(v-v_0-\Delta/3)/0.7]))$ (blue). The dotted line corresponds to the spin number density per Hz at the RASER threshold $\rho_d^{th} = 6.6 \cdot 10^{16}$/Hz. Note that all values of the cos modulated profile (red) lie below $\rho_d^{th}$, which does not mean that no RASER activity is observed.

The following three different spin density profiles in Fig. S5 serve as the input profile for the full model with interaction (Eqs.(S5-S8)): (a) The rectangular profile with a constant value of the population inversion density in the image domain $\Delta$, here $\rho_d(v) = \rho_d^{rect} = 10^{16}$/Hz (black). Additionally, a small fluctuation in $\rho_d(v)$ is considered in order to understand the contrast mechanism between regions of slightly different population inversion. (b) The cosine-modulated profile with offset, i.e. $\rho_d(v) = 5.5 \cdot 10^{15}$/Hz $(1+0.2 \cos[8\pi(v-v_0)/\Delta])$ (red) to evaluate a possible correction procedure of a distorted image.

(c) Two regions or sections separated by a 1 mm broad gap, to mimic the experimental conditions. The profile is approximated by tanh-functions, i.e. $\rho_d(v) = 4.5\cdot10^{15}$/Hz $(0.5(\tanh[(v-v_0 + \Delta/3)/0.7]$ - $\tanh[(v-v_0 +\Delta/15)/0.1]) + 0.4 (\tanh[(v-v_0 -\Delta/15)/0.1]$ - $\tanh[(v - v_0 -\Delta/3)/0.7]))$ (blue). With a chosen value for the image domain $\Delta = 10$ Hz and $w_{as} = 0.33$ Hz this corresponds to $N = \Delta/w_{as} = 30$ slices. The dotted line at $\rho_d^{th} = 6.6\cdot10^{15}$/Hz indicates the population inversion density at the RASER threshold, which is related to the threshold population inversion $d_{th} = \rho_d^{th} w = 3\cdot10^{15}$. Note that for the rectangular profile in Fig. S5 all values fulfil $\rho_d(v) > \rho_d^{th}$, while for the cosine-modulated profile $\rho_d(v) < \rho_d^{th}$. As will be shown in section 4(a) this does not mean that no RASER activity is observed for the cosine-modulated profile.

## 4(a) Simulation of 1D RASER images for a rectangular profile

The simulation of a RASER image based on a rectangular profile with constant $\rho_d(v) = \rho_d^{rect}$ has two different purposes. First, it serves as a simple model system to analyze nonlinear phenomena. Second, it serves as a reference to correct for nonlinear amplitude deformations, which occur in arbitrarily shaped RASER images.

An overall view of how the amplitude and shape of a RASER image depends on different values of the initial population inversion $d_0$ is shown in Fig. S6. In S6(A), panels I-V, five simulated RASER signals are shown for five different values $d_0 =\{5.5, 6.6, 10, 15, 30\}\cdot10^{16}$. As before, the rectangular profile is centered at $v_0 = 50$ Hz and with a constant population inversion density $\rho_d(v) = d_0/\Delta$ in the frequency range 45 Hz $< v <$ 55 Hz ($\Delta =10$ Hz). With a given step size $\delta v = w_{as} = 0.2$ Hz this results in $N = 50$ slices. The green dotted line in Fig. S6(B) indicates the threshold population density $\rho_d^{th} =d_0/\Delta = 6.6\cdot10^{1}$/Hz.

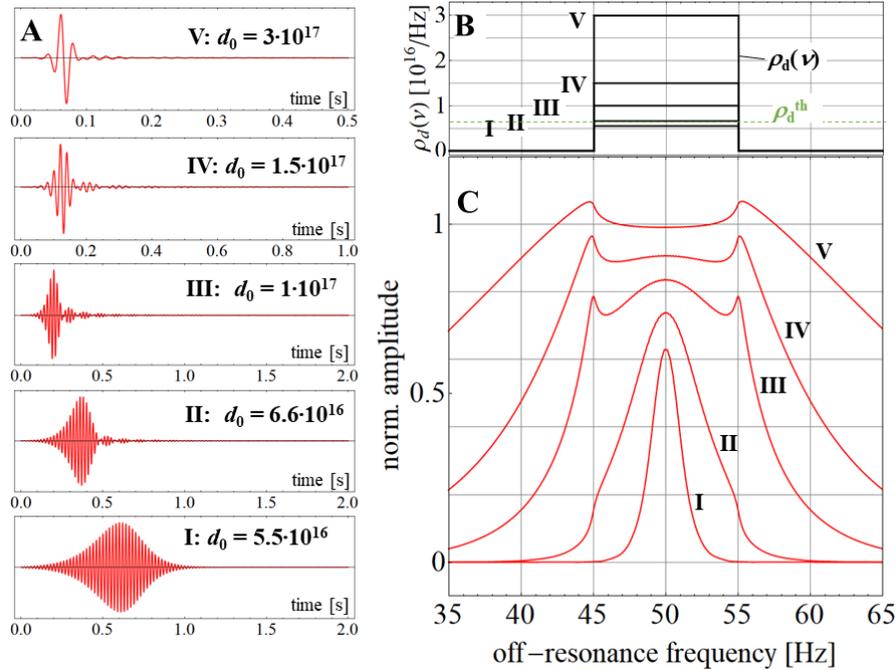

**Fig. S6. Simulated RASER images for a rectangular profile $\rho_d(v) = d_0/\Delta$.** The simulation parameters are: $T_2^* = 0.7$ s, $T_1 = 5$ s, $Q = 100$, $V_s = 0.5$ cm$^3$. Number of slices is $N = 50$, $\delta v = 0.2$ Hz, $\Delta =10$ Hz, $v_0 = 50$ Hz. (A) Simulated RASER signals for five different initial population inversions $d_0 = \{5.5, 6.6, 10, 15, 30\}\cdot10^{16}$ (panels I-V) (B) Five rectangular shaped profiles $\rho_d(v)$ (I-V, black lines). Green dotted line indicates the threshold population inversion density $\rho_d^{th} = 6.6\cdot10^{15}$/Hz. (C) RASER images obtained after Fourier transformation (absolute mode) from the RASER signals I-V in A. All images are normalized to the amplitude of image V at $v_0 = 50$ Hz. Image I is a narrow peak centered at $v = 50$ Hz although $\rho_d(v) < \rho_d^{th}$. Images II-V have decaying side lobes outside of the imaging boundaries at $v = 45$ Hz and 55 Hz.

Fig. S6(C) shows five RASER MRI images (I-V), obtained after Fourier transformation in absolute mode from five simulated RASER signals in S6(A). Let us start with image I in S6(C) ($d_0 = 5.5 \cdot 10^{16}$), where $\rho_d(v) < \rho_d^{th}$. This means that the population inversion within a slice of a width w ($\rho_d(v) \cdot w$) is smaller than the threshold population inversion $d_{th}$ required to start RASER activity. Therefore, at first glance no RASER action is expected for all $N=50$ slices. Surprisingly, a RASER burst is visible (burst I in S6(A)) and the corresponding image I in S6(C)) is a symmetrical peak with a maximum at 50 Hz and a width much narrower than $\Delta$. The reason for this shape is the cooperative action between all 50 slices. The slices close to the center at 50 Hz successfully surpass the threshold through the cooperation with their neighbors. In contrast to that, the slices on the edges (at 45 and 55 Hz) do not surpass the threshold and are not RASER active. Therefore the simulated image amplitude is maximal in the center of the image domain.

The next case for $d_0 = 6.6 \cdot 10^{16}$ is shown in image II in S6(C), where $\rho_d(v) = \rho_d^{th}$ holds. Now, all slices in the image domain $\Delta$ are RASER active and contribute to the image. Nonetheless, the image does not have a rectangular shape, but a bell shaped image with a maximum amplitude in the center is formed. Furthermore decaying sidelobes left and right from the image boundaries arise. We found that cooperative action between all slices leads to broad and complicate spectra of each slice and are responsible for the signal outside the image domain $\Delta$ (see Fig. S7).

In image III ($d_0 = 1 \cdot 10^{17}$) the amplitude at the edges is nearly as high as the amplitude in the center and the decaying sidelobes are more pronounced. In this case, the slices at the edges of the image domain benefit from the enhanced cooperative interaction between all slices. This effect is even more pronounced in the images IV ($d_0 = 1.5 \cdot 10^{17}$) and V ($d_0 = 3 \cdot 10^{17}$), with $\rho_d(v) \gg \rho_d^{th}$, where the amplitude in the center is comparable or smaller in size with respect to the amplitude at the image boundaries. The exact quantitative description and how a rectangular RASER image is formed is quite involved and will be the subject of an independent publication.

Next, we study the sensitivity of a RASER image with respect to small disturbances in the polarization distribution, in the profile $\rho_d(v)$. Therefore we compare simulated RASER images with standard imaging based on Lorentz shaped PSFs. As input profile, we assume a rectangular input profile $\rho_d(v)$ for $d_0 = 9 \cdot 10^{16}$ and for an image domain ranging from 45 Hz to 55 Hz ($\Delta = 10$ Hz, $\delta v = 0.151$ Hz, $N = 66$). On this rectangular profile, a small perturbation is added in the form of a rectangular shaped hole in the center $v_0 = 50$ Hz. The width of this hole is one half of the natural linewidth, i.e. $w/2 = 1/(2\pi T_2^*) = 0.23$ Hz, and the amplitude is 20% smaller compared to the maximum of $\rho_d(v)$. In this case, the smallest value in the hole is still above the threshold population density $\rho_d^{th} = 6.6 \cdot 10^{15}$/Hz and therefore all $N = 66$ slices are RASER active. The resulting simulated RASER image is depicted in Fig. S7(B) as I (black).

For comparison, an image based on a superposition of Lorentzian PSFs is simulated assuming a Lorentzian PSF of width w, i.e. $L(v,v^*) = 1/\left[1 + (v-v^*)^2/w^2\right]$. This PSF is folded with the profile $\rho_d(v)$ shown in Fig.7 (A) to obtain the image $S_L(v) = \int_{-\infty}^{\infty} L(v,v^*) \rho_d(v^*) dv^*$ depicted in Fig S7(B) as I (red).

Comparing both images in Fig. S7(B), there are two apparent differences in the RASER image: First, similar artifacts as shown in panel III of Fig. S6(C) for a similar population inversion arise. The amplitude of the RASER image with respect to $\rho_d(v)$ inside the imaging domain is deformed and sidelobes outside of the image domain are more pronounced. Both these artifacts are typical for a rectangular input profile above the threshold density and are not attributed to the introduced small perturbation. Second, there is a minimum in intensity in the center of the image at $v_0 = 50$ Hz. In the RASER image, the amplitude of this perturbation is about three times lower and the slope at the edges three times steeper. This simple example demonstrates the advantage and problems of RASER imaging, which is sensitive to small perturbations, but suffers from amplitude deformations inside of $\Delta$ and side lobes outside of $\Delta$.

Spectra of seven representative slices are chosen for Fig. S7(C). Each of these spectra are obtained after Fourier transform of the respective RASER signal of slice μ. The slices at the image boundaries (μ = 1 and 66) are depicted in brown, slice μ = 33 close to the center is depicted in blue, while the slices between these extremes are shown in orange (μ = 12 and 54) and green (μ = 24 and 42). The peak amplitude of the spectrum in the center (μ = 33, blue) is significantly smaller compared to slices μ = 24 and 42 (green). This motivates the increased sensitivity of RASER imaging with respect to small changes in the amplitude of $\rho_d(\nu)$. The width of each spectrum in Fig. S7(C) is close to w. Furthermore all spectra feature broad sidelobes left and right from the peak maxima, which can extend out of the image domain Δ. The highly resolved local hole with w/2 width cannot be explained by the width w for the individual $N = 66$ slices. Instead, the collective interaction between all slices generates the local image contrast. This interaction involves collective phenomena such as synchronism, line collapse and other non-linear phenomena outside of the scope of this manuscript, which focuses on the main aspects of RASER MRI and aims to keep the model as simple as possible.

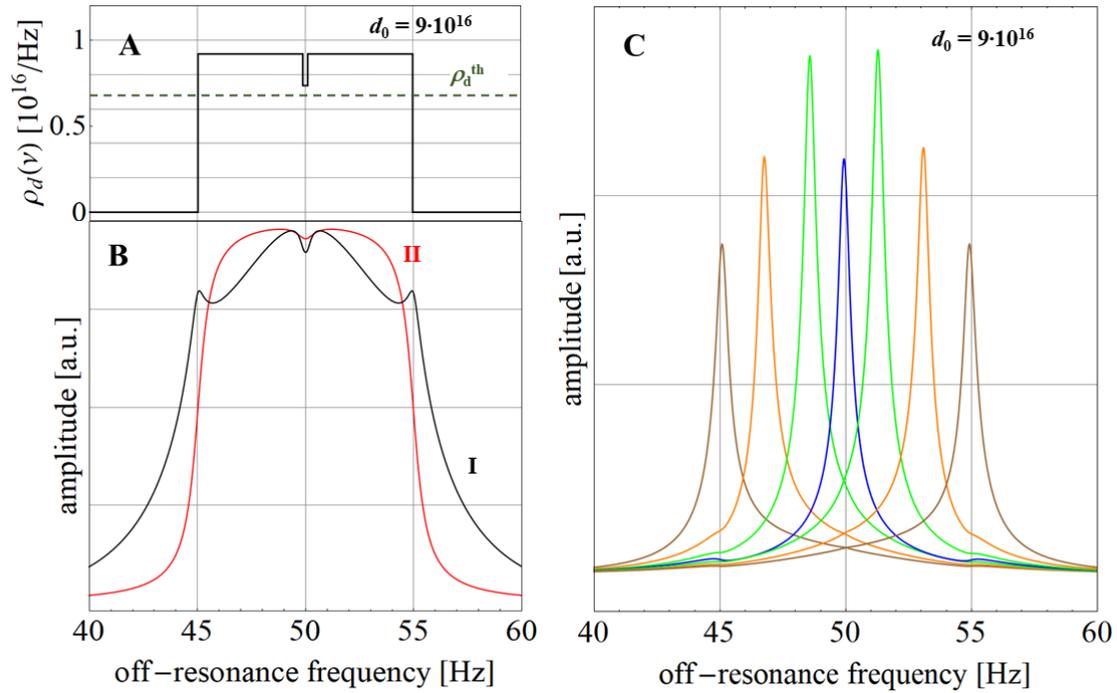

**Fig. S7. Comparison between simulated RASER images and imaging based on Lorentzian shaped PSFs.** (**A**) A rectangular input profile $\rho_d(\nu)$ is assumed with Δ = 10 Hz and $d_0 = 9 \cdot 10^{16}$ with a small rectangular hole in the center at $\nu_0 = 50$ Hz. The width of this hole is w/2 = 0.23 Hz and the depth is 20% with respect to the maximum of $\rho_d(\nu)$ and still above the threshold population density (dashed green line; $\rho_d^{th} = 6.6 \cdot 10^{15}$/Hz). (**B**): Simulated RASER image (I, black) and image based on a superposition of Lorentzian PSFs (II, red) of the profile $\rho_d(\nu)$ shown in (A). (**C**) Seven representative spectra of the 66 slices: μ = 1 and 66 (brown), 12 and 54 (orange), 24 and 42 (green) and μ = 33 (blue) obtained after Fourier transformation in the absolute mode. Note the broad shoulders left and right from the peak maxima in each of the spectra, which comes from the collective interaction between all slices. Simulation parameters for the RASER image: $N = 66$ slices, δν = 0.151 Hz, $T_2^* = 0.7$s, $T_1 = 5$ s, $Q = 100$. The width of the Lorentzian PSFs is w = $1/(\pi T_2^*)$ = 0.455 Hz.

## 4(b) Simulation of 1D RASER images with a cosine-modulated profile

In the next step we will study simulations of RASER images with more complicated profiles. One interesting example is shown in Fig. S8, depicting simulations of the time dependent RASER signals (A,D,G) and the corresponding RASER images (B,E,H) for a given offset superimposed by a cosine-modulated profile (red lines in insets of (A,D,G)).

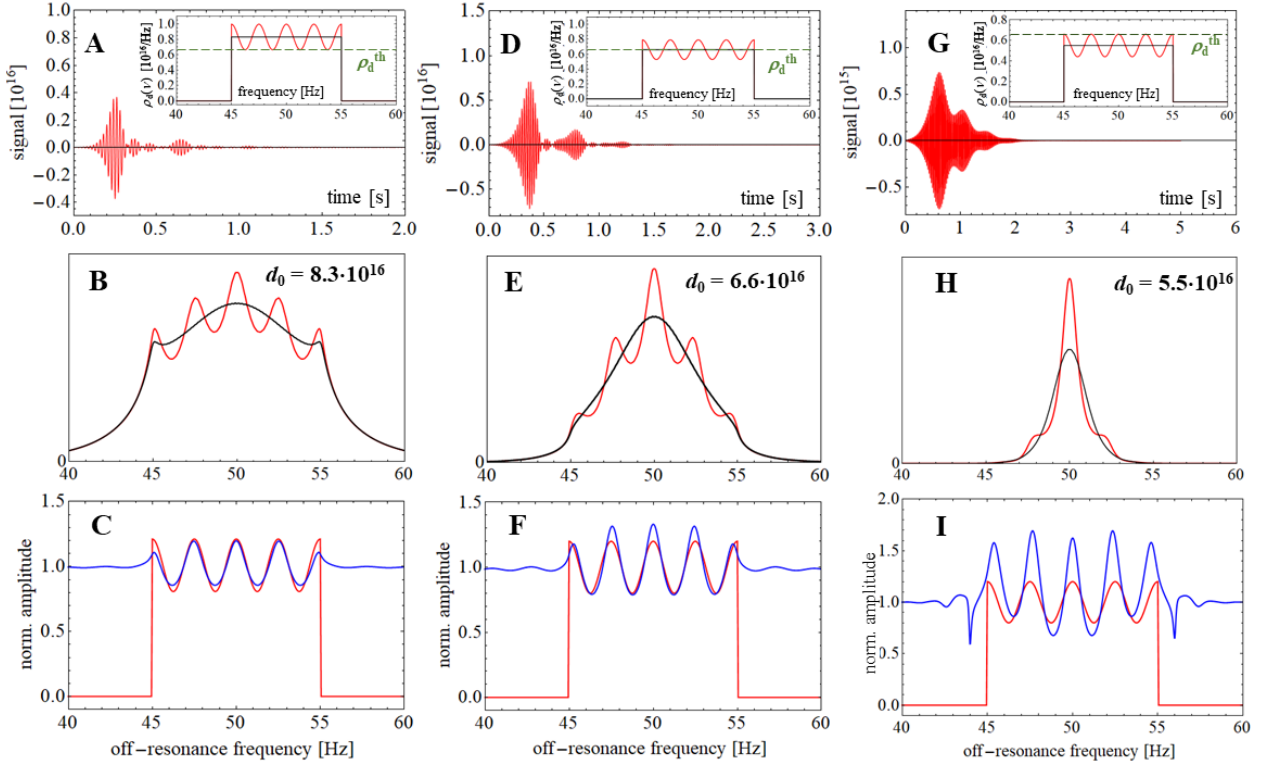

**Fig. S8. Simulated RASER images for a cosine-modulated profile with offset**. (**A, D, G**): RASER signal versus time simulated for $N = 50$ slices ($\Delta = 10$ Hz, $w_{as} = 0.2$ Hz, $v_0 = 50$ Hz, $T_1 = 5$ s, $T_2^* = 0.7$ s) and for three different offsets of the cosine-modulated profiles (insets, red), $\rho_d(v) = C_0(1 + 0.2 \cos[8\pi(v - v_0)/\Delta])$, where $C_0 = d_0/\Delta = 8.3 \cdot 10^{15}$/Hz (A), $6.6 \cdot 10^{15}$/Hz (D) and $5.5 \cdot 10^{15}$/Hz (G). The three profiles in (A,D,G) correspond to an initial population inversion of $d_0 = 8.3 \cdot 10^{16}$, $6.6 \cdot 10^{16}$, and $5.5 \cdot 10^{16}$, respectively. A black, rectangular profile in the insets of (A,D,G) serves as a reference for a global rectangular correction. The dotted green line is the density at threshold $\rho_d^{th} = 6.6 \cdot 10^{15}$/Hz. (**B,E,H**) represent the simulated RASER images for the cosine-modulated (red) and the rectangular (black) profiles for the insets of (A,D,G), respectively. (**C, F, I**): Corrected RASER images (C,F,I) are obtained through division of the amplitudes between the red and black curves in (B,E,H), respectively. Red lines correspond to the normalized spin density profiles $\rho_d^*(v) = \rho_d(v)/C_0$. The corrected image in C ($\rho_d(v) \geq \rho_d^{th}$) is in good agreement with the normalized profile $\rho_d^*(v)$. In (F,I), $\rho_d(v) \leq \rho_d^{th}$, the corrected image deviates from the normalized $\rho_d^*(v)$.

The simulation parameters and boundary conditions are given in section 5 and in the caption of Fig. S8. The threshold population density, $\rho_d^{th} = 6.6 \cdot 10^{15}$/Hz is indicated as green dashed line in the insets of Fig. S8(A,D,G). In order to resolve all the details, $N = 50$ slices are assumed for the simulations, which corresponds to an image domain of $\Delta = N \cdot w_{as} = 10$ Hz. The four periods of the profile have a modulation depth of 20%, i.e. $\rho_d(v) = C_0(1 + 0.2 \cos[8\pi(v-v_0)/\Delta])$, where the factor $C_0 = d_0/\Delta$ quantifies the average spin density. Three different offset values of $C_0$ are chosen, as indicated in the insets (A,D,G). In the first case, Fig. S8(A), $C_0 = 8.6 \cdot 10^{15}$/Hz, so $\rho_d(v) \geq \rho_d^{th}$ holds inside the whole image domain. The Fourier-transformed image in the absolute mode in S8(B), red line, reflects the four periods of the profile, and the rough envelope of the image with a maximum amplitude in the center at $v_0 = 50$ Hz is deformed relative to the profile. We know already from the previous section, that the image of a rectangular profile is deformed and characterized by a Gaussian like shape, reaching maximum amplitude at the center.

A correction of the deformed image of the cosine-modulated profile is possible by using the corresponding image of a rectangular profile. We call this procedure a global rectangular correction. The rectangular profiles are drawn as black lines in the insets of Figs. S8(A,D,G), and the constant value corresponds to the average value of the cosine-modulated profile, i.e. $\rho_d^{rect}(v) = C_0$. The black lines in (B,E,H) represent the reference images based on the corresponding three rectangular profiles in the insets of (A,D,G). The global rectangular correction procedure consists of dividing the amplitude of the cosine-modulated image by the amplitude of the reference image for each frequency. The results of this procedure are shown as blue lines in Figs. S8(C,F,I). The images of the cosine-modulated profile corrected in this manner can be directly compared to the normalized profiles (red lines in (C,F,I)), the latter being defined by $\rho_d^*(v) = \rho_d(v)/C_0$. In Fig. S8(C) there is good agreement between the corrected image (blue line) and the normalized profile (red line), in the image domain 45 Hz $< v <$ 55 Hz.

A case where half of the population inversion densities $\rho_d(v)$ are below the threshold density $\rho_d^{th}$ is shown in Figs. S8(D,E,F). The corrected image in (F) is not perfectly reflecting the normalized profile: The maxima of the corrected image are about 20% larger. An extreme case is demonstrated in Figs. S8(G,H,I), where all slices individually are below the RASER threshold $\rho_d(v) < \rho_d^{th}$. One might think that RASER activity is impossible. This would be a misconception however, due to the fact that nonlinear interaction can increase the population inversion of those slices near the maxima of $\rho_d(v)$ close to threshold. The corrected image in Fig.S8(I) retains roughly the shape of the normalized profile, but at the cost of a large deformation in the amplitude, with the minima of the corrected image being much closer to the normalized profile than the maxima.

We found that in general for small modulation depths or variations close above threshold, the images reflect the normalized profile. At higher modulation depths and further away from threshold, the nonlinear amplitude deformations increase significantly. For example, for a cosine-modulated profile with 50% modulation depth and $\rho_d(v) < \rho_d^{th}$, the image transforms into one dominant peak at the center and very small peaks close to the local maxima of $\rho_d(v)$. One open question for RASER MRI applications is whether there is an algorithm capable to correct for all nonlinear deformations.

## 4(c) Simulation of two sectors with tanh shaped edges separated by a gap

In the last example we attempt to come closer to our real experiment, in which two sections of a cylindrical sample with 8 mm inner diameter are separated by a gap of 1 mm size and pumped separately by SABRE. We determined the 1D projection experimentally with hyperpolarized high resolution Spin-Echo Imaging (see Fig.5(A), main text). These 1D images can be approximated using two step-like functions separated by the gap and the rising and falling edges of each step by tanh-functions. Examples of normalized two chamber profiles are sketched in the insets of Figs. S10(A). All five profiles correspond to an analytical expression given by the sum over four tanh step functions, i.e. $\rho_d^*(v) = A_1(\tanh[(v-v_0+b_1)/w_1]-\tanh[(v-v_0+b_2)/w_2]) + A_2(\tanh[(v-v_0-b_3)/w_3]-\tanh[(v-v_0-b_4)/w_4])$. The constants $A_1$ and $A_2$ define the maximum amplitudes of both steps, $v_0$ is the frequency offset, $b_1$, $b_3$ ($b_2$, $b_4$) denote the two positions of the rising (falling) edges of the tanh step functions relative to $v_0$, and $w_1$, $w_3$ ($w_2$, $w_4$) define the widths associated to the rising (falling) edges. The normalization of $\rho_d^*(v)$ to the value one is hereby related to the larger value of $A_1$ or $A_2$, respectively. The amplitudes of $A_1$ or $A_2$ are not necessarily equal, but depend on the pumping conditions and $T_1$ relaxation rate of each of the two chambers. The maximum value of the profile $\rho_d^*(v)$ can be compared with the normalized population inversion density at threshold, i.e. with $\rho_d^{*th} = \Delta \rho_d^{th}/d_0$.

Let us first analyze one experimental result from Fig. 5(B) main text. At $\Delta t = 8$ s, the population inversion $d_0$ has decayed significantly and all slices are below the RASER threshold. The corresponding RASER signal is shown in Fig. S9(A), while S9(B) depicts the phased spectrum and S9(C) the absolute spectrum identical with Fig. 5(B), $\Delta t = 8$ s. The applied gradient $G_z = 5.78$ mG/cm spans an image domain of $\Delta = \gamma_H G_z L = 19.5$ Hz. The signal in S9(A) is noisy and symmetrical with respect to the maximum at $t_{max} = 1.5$

s. The Fourier transformed spectrum in the phased mode (S9(B)) has the shape of a phase modulated sech function, as discussed in section 3. The corresponding absolute phased spectrum in S9(C) is a symmetrical peak centered at 127.7 Hz and the width at half height is 0.6 Hz. This width is broader compared to the Lorentzian linewidth w = 0.45 Hz. Taking into account the result for the line shape and width for single slices in Fig. S7(C) this indicates that only a few cooperating coupled slices are responsible for the shape and width of the observed spectrum.

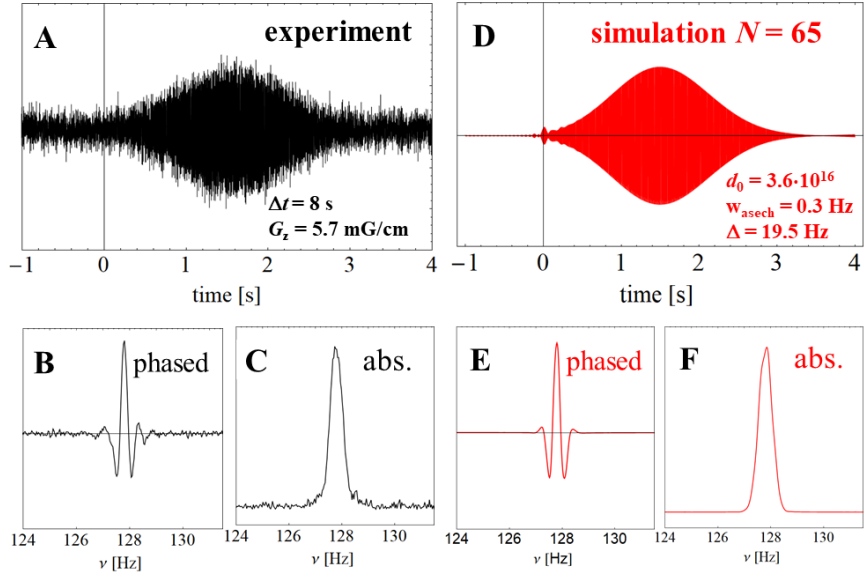

**Fig. S9. Measured (A,B,C) and simulated (D,E,F) RASER burst for an image at $\rho_d(\nu)$ below threshold.** The experiment shown in (A,B,C) is identical with Fig. 5(B), $\Delta t$ = 8 s from the main text. (**A**) Measured RASER signal acquired at $B_0$ = 7.8 mT (333 kHz $^1$H) and $G_z$ = 5.7 mG/cm with corresponding Fourier transformed spectrum in the phased (B) and in absolute mode (C). The line width at FWHM in (C) is 0.6 Hz. (**D**) Simulated RASER burst ($N$ = 65, $d_0$ =3.6·10$^{16}$, $\Delta$ = 19.5 Hz, $\nu_0$ = 132.7 Hz, $w_{as}$ = 0.3 Hz, $T_1$ = 5 s, $T_2^*$ = 0.7 s) and corresponding spectrum in the phased (E) and absolute mode (F). Note the large noise contribution in A, which indicates that RASER action is measured close to the threshold density $\rho_d^{th}$ = 6.6·10$^{15}$/Hz. All simulated signals and images based on Eqs.(1-4) in (D,E,F) are in good agreement with the measurements in (A,B,C), if a profile $\rho_d(\nu) = C_0\{0.5(\tanh[(\nu-\nu_0+6.5)/1.4]$ - $\tanh[(\nu-\nu_0+3.6)/1.4])$ + $0.25(\tanh[(\nu-\nu_0-3.6)/1.4]$- $\tanh[(\nu-\nu_0-6.5)/1.4])\}$ is assumed, where $C_0$ = 6.5·10$^{15}$/Hz and $\nu_0$ = 132.7 Hz. Although $\rho_d(\nu) < \rho_d^{th}$ the interaction between all 65 coupled slices pushes population inversion towards the maximum of $\rho_d(\nu)$ at 127.7 Hz. This leads to RASER action associated to a corresponding symmetric narrow image, which has a width of 0.6 Hz > $w_{as}$ = 0.3 Hz and a maximum value at 127.7 Hz.

In fact the observed features in Figs. S9(A,B,C) can be simulated based on the theory (Eqs. S5-S8) using an equidistant slicing of the image domain with $\delta\nu$ = 0.3 Hz. Considering the experimental image domain of $\Delta$ = 19.7 Hz, the simulations require $N$ = 65 coupled slices. For $\Delta t$ = 8 s the polarization on the right half has decayed faster compared to the left half ($A_2$ = 0.25 < $A_1$ = 0.5) and the relaxation rate on the sample walls is larger than in the bulk. Thus, the initial spin density profile is assumed as two peaks, which are each narrower than $\Delta/2$, i.e. $\rho_d(\nu)$ = 6.5·10$^{15}$/Hz {0.5($\tanh[(\nu-\nu_0+6.5)/1.4]$ - $\tanh[(\nu-\nu_0+3.6)/1.4])$ + 0.25($\tanh[(\nu-\nu_0-3.6)/1.4]$- $\tanh[(\nu-\nu_0-6.5)/1.4])$} (see Fig. S10(A), panel I).

The simulated RASER burst based on the chosen $\rho_d(\nu)$ is nearly symmetrical with respect to its maximum amplitude at $t_{max}$ = 1.5 s (see S8(D)). Both this signal and the phased and absolute spectra in S8(E,F) are in good agreement with their experimental counterpart. Note that the absolute spectrum in Fig. S9(F) is identical with Fig. 4(C), panel I.

Further away from the threshold, more slices are involved in the image and consequently the nonlinear effects are more prominent. Fig. S10(A) shows the simulated RASER signals versus time for five different values of the initial population inversion $d_0 = 2 \cdot 10^{17}$ (V), $1.5 \cdot 10^{17}$ (IV), $1.2 \cdot 10^{17}$ (III), $6.3 \cdot 10^{16}$ (II) and $3.6 \cdot 10^{16}$ (I). Further simulation parameters are given in the caption of Fig. S10. The duration of the RASER signal becomes longer with decreasing $d_0$. The corresponding normalized spin density profiles $\rho_d^*(v)$ (red lines) are depicted in the insets. For clarity normalized $\rho_d^*(v)$ are used here. The green dashed lines indicate five different values for the normalized threshold population inversion densities $\rho_d^{*th}$. In 10(A) panel I $\rho_d^{*th} = 1.06 > \rho_d^*(v)$ is assumed in the image domain $\Delta = 19.7$ Hz, thus a narrow image of the left half is expected. In S10(A) panel V, $\rho_d^{*th} = 0.4 < \rho_d^*(v)$ holds, so large nonlinear effects are expected. The five profiles $\rho_d^*(v)$ in the insets of S10(A) differ in shape and amplitude ratio between the left and right half of the phantom, which takes into account two different overall $T_1$ relaxation rates ($T_1 = 5$ s, 3 s for the left and right half, respectively) as well as locally enhanced relaxation rates at the walls of the sample chambers.

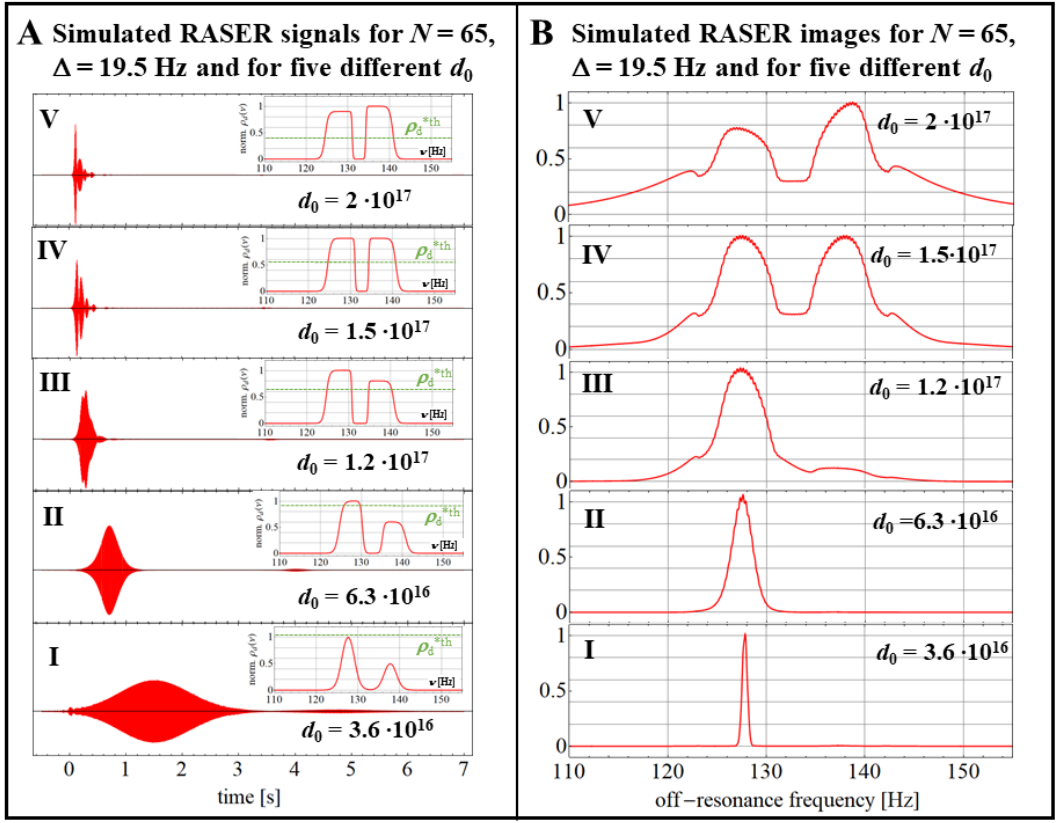

**Fig. S10. Simulation of RASER images for different $\rho_d(v)$ and $d_0$ for two chambers separated by a gap of $\Delta z = 1$ mm.** (A). RASER burst signals at the different $\rho_d(v)$ and $d_0$. Simulation parameters: $N = 65$, $\Delta = 19.5$ Hz, $w_{as} = 0.3$ Hz, $v_0 = 132.7$ Hz, $T_1 = 5$ s, $T_2^* = 0.7$ s. Insets: Corresponding normalized spin density profiles $\rho_d^*(v)$ (red lines). Green dashed lines indicate five values for the normalized threshold population inversion densities, i.e. $\rho_d^{*th} = 1.06$ (I), 0.94 (II), 0.61 (III), 0.56 (IV), and 0.4 (V). (B) The five RASER spectra I - V represent the corresponding images, obtained from the RASER signals in (A) after Fourier transformation in the absolute mode. The images reflect roughly the features of the experimentally measured RASER images (Fig. 5B, main text), including the side lobes in images IV,V appearing outside the image boundaries at 122.95 Hz and 142.45 Hz. For convenience the normalized population inversion density $\rho_d^*(v)$ is introduced in the insets of Fig. S10(A), where the maximum of the profile $\rho_d^*(v)$ is normalized to one. The corresponding normalized threshold population density $\rho_d^{*th}$ is indicated by the dotted green lines.

The five RASER spectra I–V in Fig. S10(B) represent the corresponding images, which are obtained from the RASER signals in S10(A) after Fourier transformation in the absolute mode. With increasing $d_0$, the images change from (I) a narrow peak, to (II) a broader image centered at 127.7 Hz, (III) an image where the right half starts to be visible, (IV) an image where both halves with equal amplitudes separated by the gap appear and finally (V) to an image where the right half is larger in amplitude. The simulated image I in S10(B) is identical to Fig. S9(F) and is discussed above. The other four simulated images II-V roughly reflect the features of the experimentally measured RASER images in Fig. 5B, main text, including the side lobes in images IV,V appearing outside the image boundaries and non-zero values in the gap.

There are two additional phenomena, sometimes observed in the measured images: Pronounced ripples and regions with strongly changing amplitudes. For example ripples can be seen in Fig. 5(B), main text, at $\Delta t = 1$ s, 4 s and 5 s at positions 1 mm < z < 2 mm. An example for peaks with strongly changing amplitudes is shown in Fig. 5(B), main text, for $\Delta t = 1$ s at position z = – 1 and 3 mm. Possible imaging artifacts for RASER MRI are discussed in the next section.

## 4. 2D RASER image artifacts and 1D projections

This section focuses on the artifacts that can arise in 1D and 2D RASER images. The manuscript mainly discusses 1D images, also called projections, but in Fig. 4(A) and (B) of the main text two different 2D images are shown. A spin echo image for reference and a RASER image. These 2D images are generated using projection reconstruction of 30 1D images measured from 30 angular directions. In these 2D images, artifacts arise due to the projection reconstruction algorithm as well as within the 1D projections themselves.

Projection reconstruction generates star artifacts. These are well known, as projection reconstruction is widely used e.g. in imaging methods such as computed tomography (CT). The star artifacts are more pronounced further away from the center. This can for example be seen in the bottom left corner of Fig. S11(A) and the top left and right corner of Fig. S11(B). Star artifacts can be reduced and the angular resolution increased by measuring more projections.

Other artifacts and features in the 2D images stem from the $p$-$H_2$ delivery system, necessary for SABRE pumping. The $p$-$H_2$ is introduced through a capillary in each of the chambers. In the SEI, they can be seen as dark spots each corresponding to the location of a capillary for $p$-$H_2$ supply. For RASER MRI, the capillaries can additionally result in nonlinear distortions of a given projection, as RASER MRI is very sensitive to local fluctuations in polarization. Additionally, the $p$-$H_2$ bubbling introduces a motion of the liquid in both chambers. This motion stops after about 1-2 s. To ensure a motion-free reference image, $\Delta t$ = 5s is chosen for SEI. The RASER image however, is very sensitive to the initial population inversion as visualized in Fig. 5 of the main text. To ensure that both chambers have a population inversion in a regime where an image is formed, $\Delta t = 2$s is chosen for RASER MRI. This leaves the RASER image more susceptible to small residual motion.

Most artifacts can be identified in the 1D projections. Thus, for the 2D RASER image in Fig. 4(B) of the main text, five projections (I-V) are selected. Their gradient directions are drawn as colored lines in Fig. S11(B). The 1D images of these chosen angles are depicted in Fig. S11(C). When choosing a gradient direction perpendicular to the gap between the half circles of the phantom, the gap can be identified as a minimum amplitude in the middle of the projection. Here, projections III (green) and IV(blue) are close to this condition. For the 1D images in Fig. 3 of the main text, the gradient is chosen perpendicular to the gap as discussed there. A gradient parallel to the direction of the gap yields a projection without a minimum amplitude in the middle, similar to the projection of a circle (see Fig. S11(C), projection V, orange).

The most prominent artifacts in the 2D RASER image are interference lines through the entire image. They can be identified within the 1D projections that have a gradient direction perpendicular to the observed interference line. In the projection they can be identified as "spikes" at the given position. This is visualized exemplarily for two interference lines, marked by stars and arrows in the 2D image. They stem from projections II (red) and III (green), respectively. These artifacts can also be identified in the projections and are encircled and marked with stars in Fig. S11(C). One possible reason for such interference artifacts is the sensitivity of the coupled RASER modes to local disturbances. Disturbances can be caused by residual motion as described above, the capillaries for $p$-$H_2$ delivery and fluctuations in the profile $\rho_d(v)$, produced by the SABRE pumping.

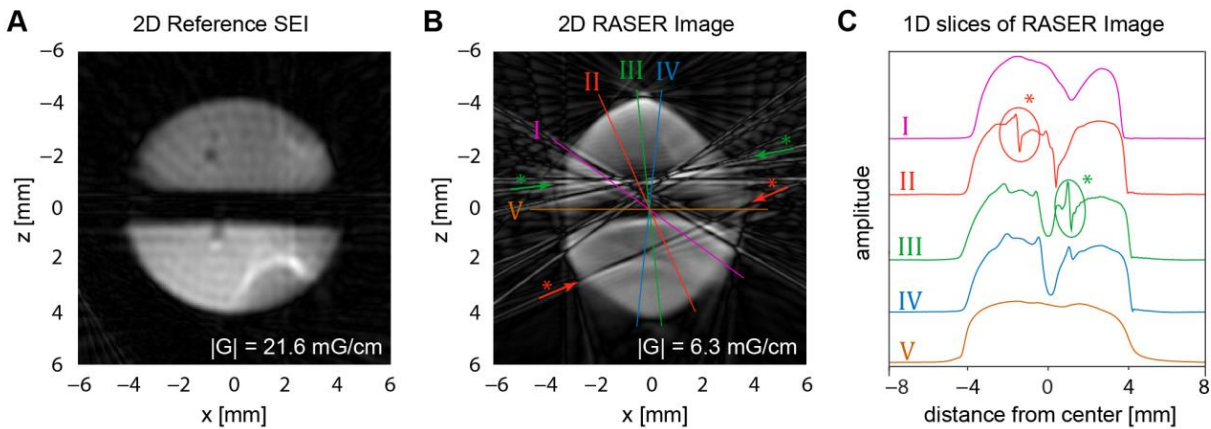

**Fig. S11. 2D SEI and 2D RASER image from the main text (Fig. 4) and selected RASER imaging projections.** (**A**) 2D- SEI, (**B**) 2D RASER MRI and (**C**) selected RASER image projections measured at 3.9 mT. To obtain (**A**) and (**B**), 30 projections are measured with the sequence in Fig.2(C, D), each from different angles by varying $G_x$ and $G_z$ such that $G_x^2+G_z^2$ = const. The 2D images are obtained after projection reconstruction. In (**A**), the two capillaries used for $p$-$H_2$ supply are visible around x = –1 mm, z = 0.5 mm and x = –1.5 mm, z = –2 mm for each chamber. In (**C**), five 1D projections (I-V) used to reconstruct the 2D RASER image are displayed. The direction of the gradient for each of the five projections is depicted as a colored line (I-V) in (**B**). The RASER image (**B**) is recorded at a 3.5 times smaller gradient than (**A**), but both spatial resolutions are similar. The RASER image is plagued by interference lines. Two of these lines are marked in both (**B**) and (**C**) by stars corresponding to projection II (red) and III (green), respectively. The direction of the interference line in the 2D RASER image in (**B**) is perpendicular to the gradient direction (marked by arrows). These two artifacts in (**B**) can be identified as spikes in the corresponding 1D projections (II) and (III) in (**C**) and are highlighted by a star. The origin of these artifacts is discussed in the text.

Further artifacts that arise in a 1D RASER image are leaking signal into a gap as well as sidelobes that arise outside of the image domain. They are small in the 2D images depicted here, but can play a major role in 1D RASER images recorded at other experimental conditions. These leaking and sidelobe artifacts are discussed extensively in section 4a for a rectangular profile $\rho_d(v)$.

Other phenomena such a global and relative dipolar shifts might contribute to the observed artefacts. Past studies showed that global dipolar shift effects are quite small for SABRE pumped samples. For example the total $^1$H dipolar shift generated by the SABRE pumped liquid, as described in the supplement of (*19*), was in the order of 200 mHz ($B_{dip}$ ~5 nT). This global dipolar field decays with time as the population inversion $d_0$ is depleted during the RASER burst. This time dependent global dipolar field may alter the shape of the image by up to 0.2 Hz, which is in the order of the width of one slice $\delta v = w_{as}$. Whether these time dependent dipolar shifts contribute to the interference artifacts or could produce chaotic regions is still an open question. To keep the RASER MRI model as simple as possible, we also neglected dipolar contributions in the simulations.

# 5. Conclusion & Outline

In conclusion we have presented the theory of RASER MRI, which in the imaging domain $\Delta$ can be approximated by a network of $N = \Delta/\delta v$ equidistant nonlinear coupled slices (Eqs.(S5-S8)). Various nonlinear effects, such as high sensitivity to local variations in the input profile $\rho_d(v)$, amplitude deformations and edge artefacts, are predicted by the theory and indeed have been observed in various experiments.

Interestingly, the mathematical form of Eqs.(S5-S8) is closely related to many descriptions of nonlinear and collective phenomena in other fields of Science. Especially self-organized processes, the topic of synergetics uses order parameters and adiabatic elimination of fast variables to derive the LASER equations. Looking at the multimode case, H. Haken has shown a close relationship between the LASER assuming a continuous number of modes with the Ginzburg-Landau theory of superconductivity (*15*). A similar correspondence between uniformly distributed oscillators with non-local coupling with the Ginzburg-Landau theory of superconductivity is shown in the article by Y. Kuramoto and D. Battogktokh (*33*). In fact, the Kuramoto model of synchronized oscillators, which is equivalent to Eq.(S7), is a subset of our model of RASER MRI. As outlined in the book from S. Strogatz (*14*) and in many articles (*33, 34, 51*) phase synchronism in non-linear coupled oscillators occur in many very different fields in physics and biology. Examples are biological rhythms, synchronized action of fireflies and tree frogs, electrically coupled Josephson arrays (*14*), mechanically coupled metronomes (chimera states) (*52*) spin torque nano-oscillators (*47*) and synchronized spin-valve oscillators (*53*), and the dynamics of neural oscillator networks (*48*).

Furthermore Hakens equations for one single LASER mode without applying the enslaving principle are apart from a variable transformation identical to the pioneering Lorenz equations (*14, 54*), the ladder of which describes chaos in a three dimensional space (*15*). In a similar manner it has been shown that chaos arises for the case of two enslaved RASER modes (*16*), which involves the evolution of four independent parameters $d_1$, $\alpha_1$, $d_2$, and $\alpha_2$. At present the exact analysis for why and how chaos arises in $N$ coupled RASER modes is unknown.

Due to the strong links of Eqs.(S5-S8) to many different fields we believe that the presented theory of RASER MRI may be a base for a deeper understanding of self-organizing processes based on both adiabatic elimination of fast variables and on synchronism.